\documentclass[article]{IEEEtran}

\usepackage[utf8]{inputenc} 
\usepackage[T1]{fontenc}
\usepackage{url}
\usepackage{ifthen}
\usepackage{cite}
\usepackage[cmex10]{amsmath} 
\usepackage{mathtools}
\usepackage{amssymb}
\usepackage{breqn}

\usepackage{balance}
\usepackage{enumitem}

\usepackage{graphicx}
\usepackage{subcaption}

\newtheorem{definition}{Definition}
\newtheorem{lemma}{Lemma}[section]
\newtheorem{theorem}{Theorem}[section]

\newcommand{\Prob}{\mathsf{P}}
\newcommand{\Expect}{\mathsf{E}}

\long\def\symbolfootnote[#1]#2{\begingroup%
\def\thefootnote{\fnsymbol{footnote}}\footnote[#1]{#2}\endgroup}

\begin{document}
\title{A Bayesian Theory of Change Detection in Statistically Periodic Random Processes} 

\author{Taposh Banerjee, \textit{Member, IEEE\vspace{-3ex}}, Prudhvi Gurram, and Gene Whipps
}
%

\maketitle

\symbolfootnote[0]{\small
The work of Taposh Banerjee was supported
by a grant from the U.S. Army Research Laboratory, W911NF1820295.

Taposh Banerjee is with the Department of Electrical and Computer Engineering, 
The University of Texas at San Antonio, San Antonio, TX, 78249. Prudhvi Gurram is with Booz Allen Hamilton, McLean, VA 22102, 
and the U.S. Army Research Laboratory, Adelphi, MD 20783. Gene Whipps is with the U.S. Army Research Laboratory, Adelphi, MD 20783
(e-mail: taposh.banerjee@utsa.edu; gurram_prudhvi@bah.com; gene.t.whipps.civ@mail.mil). 

Copyright (c) 2019 IEEE. Personal use of this material is permitted.  However, permission to use this material for any other purposes must be obtained from the IEEE by sending a request to pubs-permissions@ieee.org.
}

\begin{abstract}
A new class of stochastic processes called independent and periodically identically distributed (i.p.i.d.) processes is defined
to capture periodically varying statistical behavior. A novel Bayesian theory is developed for detecting a change in the distribution of 
an i.p.i.d. process. It is shown that the Bayesian change point problem can be expressed as a problem of optimal control of a Markov decision process (MDP)
with periodic transition and cost structures. 
Optimal control theory is developed for periodic MDPs for discounted and undiscounted total cost criteria. A fixed-point equation is obtained that is satisfied by the optimal cost function. 
It is shown that the optimal policy for the MDP is nonstationary but periodic in nature. A value iteration algorithm is obtained to compute the optimal cost function. 
The results from the MDP theory are then applied to detect changes in i.p.i.d. processes. It is shown that while the optimal change point algorithm is a stopping 
rule based on a periodic sequence of thresholds, a single-threshold policy is asymptotically optimal, as the probability of false alarm goes to zero. 
Numerical results are provided to demonstrate that the asymptotically optimal policy is not strictly optimal. 
\end{abstract}


\section{Introduction}
In the problem of quickest change detection, the objective is to detect a change in the distribution of a sequence of random 
variables with the minimum possible delay, subject to a constraint on the rate of false alarms \cite{veer-bane-elsevierbook-2013, poor-hadj-qcd-book-2009, tart-niki-bass-2014}. 
Optimal or asymptotically optimal algorithms for quickest change detection are available in the literature. The results can be divided broadly into two categories: results for independent and identically distributed (i.i.d.) processes with algorithms that can be computed recursively and using finite memory and enjoy strong optimality properties 
\cite{mous-astat-1986, shir-opt-stop-book-1978}, and results for non-i.i.d. data with algorithms that cannot be necessarily computed recursively or using finite memory but are asymptotically optimal \cite{lai-ieeetit-1998, tart-veer-siamtpa-2005, tartakovsky2017asymptotic, Pergamenchtchikov2018}. 

In this paper, we develop theory and algorithms for detecting changes in stochastic processes that have
periodically varying statistical characteristics. In this non-i.i.d. setting, 
we will show that the optimal algorithms can be computed recursively and using finite memory. 
The motivation for this problem comes from the following anomaly detection problems in cyber-physical systems and biology where 
such periodic behavior is observed. 
\begin{enumerate}[leftmargin=*]
\item Traffic monitoring: In \cite{bane-fusion-2018} and \cite{bane-globalsip-2018}, we reported results on multimodal traffic data we collected from NYC around a 5K run; during, before 
and after the run. 
We collected CCTV images, Twitter and Instagram data. We extracted counts of persons and vehicles appearing in CCTV images over time using a deep neural network-based object 
detector. We observed that in the absence of the event (in the normal regime), the counts have a periodic statistical behavior (over a day or a week) with increased intensity every day during morning and evening rush hours.  
\item Social networks: We also observed in \cite{bane-fusion-2018} and \cite{bane-globalsip-2018} that the aggregate social network behavior shows periodic characteristics 
under the normal regime. Also, the total number of Instagram messages posted in a rectangular area around the CCTV cameras showed periodic behavior. 
\item Power grid monitoring: The power usage by end users have a periodic pattern with low usage at nighttime and high usage at daytime \cite{chen2016quickest}.
\item Neural spike patterns: In brain-computer interface studies where single neural spike data is collected, the spike firing pattern can exhibit statistically periodic behavior 
in the absence of any external stimuli; see, for example, \cite{bane-NER-2019}. 
\item ECG: Several biological signals including the ECG have a periodic behavior \cite{PhysioNet}.
\end{enumerate}
The problem of anomaly detection in the above-mentioned applications can be seen as a problem of detecting changes in deviation from periodic statistical behavior. 

In this paper, we develop a Bayesian theory for anomaly detection in problems where the statistical characteristics are periodic. We introduce a class of stochastic processes called independent and periodically identically distributed (i.p.i.d) processes that can be used to model such periodic statistical behavior. We then develop algorithms
for quickest detection of changes in i.p.i.d. processes and prove their optimality with respect to the Bayesian criterion of Shiryaev \cite{shir-siamtpa-1963}. 
In the Shiryaev formulation, the objective is to detect a change in the distribution of a stochastic process to minimize the average detection delay, 
subject to a constraint on the probability of false alarm. In the Shiryaev problem, each time we take an observation, we pay a penalty for 
delay if the change has already occurred. If an alarm is raised and the change has not yet occurred, we pay a penalty for a false alarm. 
In this paper, we also study a more general or modified Shiryaev formulation where the penalties on the delay and the false alarm are dependent on time. The latter problem is relevant 
for detecting changes in non-i.i.d. processes. 
The definition of i.p.i.d. processes and precise problem formulations are given in Section~\ref{sec:model}. 

When the observations are i.i.d. before and after the change, optimal algorithms are obtained in the Bayesian setting of Shiryaev using optimal stopping theory or dynamic programming \cite{shir-opt-stop-book-1978}, \cite{bert-dyn-prog-book-2017}, \cite{krishnamurthy2016partially}. However, since the processes under investigation here are not i.i.d., but i.p.i.d., the traditional optimal control theory cannot be applied. We show in this paper that the change detection problems for i.p.i.d. processes we study in this paper can be 
mapped to a problem of optimal control of a Markov decision process (MDP) with a nonstationary but periodic transition and cost structure.
As a result, in Section~\ref{sec:MDP}, we first develop an optimal control theory 
for periodic MDPs. With a view towards optimal stopping, we develop the optimal control 
theory for the total cost problem with finite control spaces \cite{bert-dyn-prog-book-2017}. 
For stationary problems, the optimal policy can be obtained using the framework of dynamic programming 
and can be shown to be Markovian and stationary \cite{bert-dyn-prog-book-2017}. 
A general recipe for solving nonstationary (including periodic) problems can also be found in \cite{bert-dyn-prog-book-2017}; e.g., see pg. 256.
In fact, it is suggested in \cite{bert-dyn-prog-book-2017} that the optimal policy for periodic problems is periodic in nature; see also \cite{martin1967optimal}. 
In this paper, we explicitly derive 
optimal policies for periodic problems using a more direct approach and prove that the optimal policies are indeed periodic. We obtain a fixed point equation
satisfied by the optimal cost function. We also obtain a value iteration algorithm for computing the optimal cost using which the periodic optimal policy can be computed. 
The optimal control theory developed here for periodic MDPs should be of independent interest for other control applications as well. 

In Section~\ref{sec:ApplnOST}, we apply the optimal control theory developed in Section~\ref{sec:MDP} to the change detection problems. We show that
the optimal change detection algorithm is periodic. 
The change detection statistic, as in the classical i.i.d. setting, is the \textit{a posteriori} probability 
that the change has already occurred. But, unlike the i.i.d. setting where a single-threshold policy is strictly optimal, the stopping threshold for the i.p.i.d. problem varies with time. In fact, we show that the sequence of thresholds is periodic. 
We provide examples where a periodic and nonstationary policy is strictly better than a single threshold stationary policy. 

In Section~\ref{sec:Asymptotic}, however, we show that if the constraint on the probability of false alarm is small, then we can, in fact, use a fixed time-invariant threshold. 
Specifically, we show that a single-threshold algorithm is asymptotically optimal for the classical Shiryaev formulation by showing that the proposed algorithm achieves a universal lower bound on the delay of any change detection procedure \cite{tart-veer-siamtpa-2005}. We will show that while the exact optimality result and periodic MDP theory is valid only for geometric priors, the asymptotic optimality result is valid for a large class of distributions on the change point.


\section{Mathematical Model}\label{sec:model}
We begin by first introducing a class of stochastic processes that can be used to model data showing periodic statistical behavior. 
\begin{definition}
A stochastic process $\{Y_n\}$
is called independent and periodically identically distributed (i.p.i.d) if 
\begin{enumerate}[leftmargin=*]
\item The random variables $\{Y_n\}$ are independent.
\item If $Y_n$ has density $f_n$, for $n \geq 1$, then there is a positive integer $T$ such 
that the sequence of densities $\{f_n\}$ is periodic with period $T$:
$$
f_{n+T} = f_n, \quad \forall n \geq 1. 
$$
\end{enumerate}
\end{definition}
We say that the process is i.p.i.d. with the law $(f_1, \cdots, f_T)$. Note that the law of an i.p.i.d. process is completely characterized by the finite-dimensional product distribution involving $(f_1, \cdots, f_T)$. 
We assume that in the normal regime, the data can be modeled as an i.p.i.d. process. At some point in time, due to an event, 
the distribution of the i.p.i.d. process deviates from $(f_1, \cdots, f_T)$. Our objective in this paper is to develop algorithms
to observe $\{Y_n\}$ in real time and detect changes in the distribution as quickly as possible, subject to a constraint 
on the rate of false alarms. If the period $T=1$, an i.p.i.d. process reduces 
to an i.i.d. process. Optimal algorithms for change detection in i.i.d. processes has been extensively developed in the 
literature \cite{veer-bane-elsevierbook-2013, poor-hadj-qcd-book-2009, tart-niki-bass-2014, mous-astat-1986, shir-opt-stop-book-1978, lai-ieeetit-1998, tart-veer-siamtpa-2005, tartakovsky2017asymptotic, Pergamenchtchikov2018}. 

We now define a change point model. Consider another periodic sequence of densities $\{g_n\}$ such that 
$$
g_{n+T} = g_{n}, \quad \forall n \geq 1. 
$$
We assume that at some point in time $\nu$, 
called the change point in the following, the law of the i.p.i.d. process is governed not by the densities $(f_1, \cdots, f_T)$, 
but by the new set of densities $(g_1, \cdots, g_T)$:
\begin{equation}\label{eq:changepointmodel}
Y_n \sim 
\begin{cases}
f_n, &\quad \forall n < \nu, \\
g_n &\quad \forall n \geq \nu.
\end{cases}
\end{equation}
The densities $(g_1, \cdots, g_T)$ need not be all different from the set of densities $(f_1, \cdots, f_T)$, 
but we assume that there exists at least an $i$ such that they are different:
\begin{equation}\label{eq:diffpdfassum}
g_i \neq f_i, \quad \text{for some } i = 1, 2, \cdots, T. 
\end{equation}

\subsection{Classical Shiryaev Formulation}

Let $\tau$ be a stopping time for the process $\{Y_n\}$, i.e., a positive integer-valued random variable such that the event $\{\tau \leq n\}$ belongs
to the $\sigma$-algebra generated by $Y_1, \cdots, Y_n$. In other words, whether or not $\tau \leq n$ is completely determined by the first $n$ 
observations. We declare that a change has occurred at the stopping time $\tau$. To find the best stopping rule to detect the change in distribution, we
need a performance criterion. Towards this end, we model the change point $\nu$ as a random variable with a prior $\pi$:
$$
\pi_n = \Prob(\nu =n), \quad \text{ for } n = 1, 2, \cdots.
$$
For each $n \in \mathbb{N}$, we use $\Prob_n$ to denote the law of the observation process $\{Y_n\}$ when the change occurs at $\nu=n$, and use $\Expect_n$ to denote the corresponding expectation. 
Using this notation, we define the average probability measure
$$
\Prob^\pi = \sum_{n=1}^\infty \pi_n \; \Prob_n.
$$
To capture a penalty on the false alarms, we use the probability of false alarm defined as 
$$
\Prob^\pi(\tau < \nu).
$$
To penalize the detection delay, we use the average detection delay given by
$$
\Expect^\pi\left[(\tau - \nu)^+\right]
$$
or its conditional version 
$$
\Expect^\pi\left[ \tau - \nu | \tau \geq \nu \right],
$$
where $x^+ = \max\{x, 0\}$.

The optimization problem we are interested in solving is
\begin{equation}\label{eq:ShiryProb}
\min_{\tau \in \mathbf{C}_\alpha} \Expect^\pi\left[(\tau - \nu)^+\right],
\end{equation}
where 
$$
\mathbf{C}_\alpha = \{\tau: \Prob^\pi(\tau < \nu) \leq \alpha\},
$$
and $\alpha$ is a given constraint on the probability of false alarm. In the above problem, we can also 
use the conditional version of the delay $\Expect^\pi\left[ \tau - \nu | \tau \geq \nu \right]$. 

When the change point $\nu$ is a geometric random variable, the classical approach to solving problem \eqref{eq:ShiryProb} is to 
solve a relaxed version using dynamic programming. Specifically, let 
$$
\Prob(\nu =n) =  (1-\rho)^{n-1} \rho, \quad \text{ for } n = 1, 2, \cdots.
$$
Then
$$
\Prob^\pi = \sum_{n=1}^\infty (1-\rho)^{n-1} \rho \; \Prob_n.
$$
The relaxed Bayesian optimization problem is
\begin{equation}\label{eq:relaxedShiryProb}
\min_{\tau} \; \Expect^\pi \left[( \tau - \nu)^+\right] + \lambda_f \; \Prob^\pi(\tau < \nu),
\end{equation}
where $\lambda_f > 0$ is a penalty on the cost of false alarms. 
The above optimization problem can be stated as a problem in partially observable MDPs (POMDPs); see Section~\ref{sec:ModShir}, and also \cite{veer-bane-elsevierbook-2013, krishnamurthy2016partially}, and \cite{bane-fusion-2018}. Specifically, define $p_0=0$ and 
\begin{equation}\label{eq:Shirpn}
p_n = \Prob^\pi(\nu \leq n | Y_1, \cdots, Y_n), \text{ for } n \geq 1. 
\end{equation}
Then, it can be shown that problem \eqref{eq:relaxedShiryProb} is equivalent to solving
\begin{equation}\label{eq:ShirPOMDP}
\min_{\tau} \; \Expect^\pi\left[ \sum_{n=0}^{\tau-1} p_n + \lambda_f (1-p_\tau)\right].
\end{equation}
If the period $T=1$ and the processes are i.i.d., then the problem in \eqref{eq:ShirPOMDP} can be solved using 
classical belief state MDPs \cite{veer-bane-elsevierbook-2013, krishnamurthy2016partially, bane-fusion-2018, bert-dyn-prog-book-2017}. 
However, in our case, the belief updates are not stationary. 

\medskip
\begin{lemma}\label{lem:ShirRecPn}
The belief $p_n$ in \eqref{eq:Shirpn} can be recursively computed using the following equations: $p_0=0$ and for $n \geq 1$, 
\begin{equation}\label{eq:beliefUpdates}
p_n = \frac{\tilde{p}_{n-1} \; g_n(Y_n)}{\tilde{p}_{n-1} \; g_n(Y_n) + (1-\tilde{p}_{n-1}) f_n(Y_n)},
\end{equation}
where 
$$
\tilde{p}_{n-1} = p_{n-1} + (1-p_{n-1}) \rho. 
$$
\end{lemma}
\begin{IEEEproof}
The proof is provided in the appendix. 
\end{IEEEproof}
\medskip

Note that the likelihood ratios are a function of the time index $n$. Thus, the belief updates are nonstationary. However, because
the processes are i.p.i.d. in nature and there are only finitely many densities $(f_1, \cdots, f_T)$ and $(g_1, \cdots, g_T)$, 
the belief updates have a periodic structure that repeats after $T$ time slots.

The optimal stopping problem \eqref{eq:ShirPOMDP} cannot be solved using classical optimal stopping theory or dynamic programming \cite{shir-opt-stop-book-1978}, \cite{bert-dyn-prog-book-2017}, \cite{krishnamurthy2016partially}. This is because in these theories it is assumed that the Markov process to be controlled is homogeneous. 
The Markov process $\{p_n\}$ to be controlled in \eqref{eq:ShirPOMDP}  is not homogeneous. However, as shown in Lemma~\ref{lem:recursion}, the transition structure is periodic. 
Motivated by this observation, in Section~\ref{sec:MDP}, we develop the optimal stopping theory or optimal control theory for periodic MDPs. In Section~\ref{sec:ApplnOST}, we will apply 
the results obtained for periodic MDPs to the periodic optimal stopping problem \eqref{eq:ShirPOMDP}.

\subsection{Modified Shiryaev Formulation}\label{sec:ModShir}
In this section, we formulate a more general optimal stopping problem than stated in \eqref{eq:ShirPOMDP}. In the problem in \eqref{eq:ShirPOMDP}, 
the delay penalty at all times is $1$ unit and the false alarm penalty is $\lambda_f$ units. Since the processes under study here are not i.i.d., 
we formulate a POMDP where the delay and false alarm penalties are a function of time. Since we are investigating i.p.i.d. processes, we assume
that the delay and false alarm penalties are periodic as well. The precise problem formulation is given below.

\begin{itemize}
\item \textit{States}: 
Let $\{\Theta_k\}_{k \geq 0}$ be a sequence of states with values $\{\theta_k\}$. The state process is a finite-state Markov chain taking values
\begin{equation}
\Theta_k \in \{\mathcal{A}, 0, 1\}, \; \forall k.
\end{equation}
The state $\mathcal{A}$ is a special absorbing state introduced for mathematical convenience in a stopping time POMDP \cite{krishnamurthy2016partially}. 

\medskip
\item \textit{Control}: The control sequence for the POMDP is the process $\{U_k\}_{k \geq 0}$ taking values $\{u_k\}$ and is binary valued: 
\begin{equation}
U_k \in \{1 \;(\text{stop}), \; 2 \;(\text{continue}) \}.
\end{equation} 
The control $U_k=2$ is used to continue the observation process and $U_k=1$ is used to stop it. 
At the time of stopping, an alarm is raised indicating that a change in the distribution of the observations has occurred. 
\medskip
\item \textit{Observations}: The distribution of the observations $\{Y_k\}$ depends on the state $\Theta_k$ and if the control is to continue: for $k \geq 1$,
\begin{equation}
\begin{split}
(Y_k | \Theta_k = 0, &\; U_{k-1}=2) \sim f_k \\
(Y_k | \Theta_k = 1, &\; U_{k-1}=2) \sim g_k,
\end{split}
\end{equation}
with the understanding that the observation process is i.p.i.d. with law $(f_1, \cdots, f_T)$ before the change, 
and i.p.i.d. with law $(g_1, \cdots, g_T)$ after the change. No observation is collected at time $k=0$ and 
the distribution of the observations when the state equals $\mathcal{A}$ is irrelevant to the problem.
%
\medskip
\item \textit{Transition Structure}: The Markov chain $\{\Theta_k\}$ evolves according to a transition structure that depends on the control process $\{U_k\}$. Let $P(u_k)$ be the transition matrix for the 
Markov chain, given the control is $U_k=u_k$. Then, we have
\begin{equation}
P(u_k)= \begin{cases}
               P_1, \quad \text{if } u_k = 1\\
               P_2, \quad \text{if } u_k = 2,
            \end{cases}
\end{equation}
where 
\begin{equation}
P_1 = \begin{bmatrix}
p_{\mathcal{A}\mathcal{A}}&p_{\mathcal{A}0} &p_{\mathcal{A}1}\\
p_{0\mathcal{A}}&p_{00} &p_{01}\\
p_{1\mathcal{A}}&p_{10} &p_{11}\\
\end{bmatrix}
=
\begin{bmatrix}
    1 &0 & 0\\
    1 &0 & 0\\
    1 &0 & 0\\ 
\end{bmatrix}
\end{equation}
and
\begin{equation}
P_2 = \begin{bmatrix}
p_{\mathcal{A}\mathcal{A}}&p_{\mathcal{A}0} &p_{\mathcal{A}1}\\
p_{0\mathcal{A}}&p_{00} &p_{01}\\
p_{1\mathcal{A}}&p_{10} &p_{11}\\
\end{bmatrix}
=
\begin{bmatrix}
    1 &0 & 0\\
    0 &1-\rho & \rho\\
    0 &0 & 1
\end{bmatrix}.
\end{equation}
The initial distribution $\tilde{\pi}_0$ for the Markov chain $\{\Theta_k\}$ is
\begin{equation}
\tilde{\pi}_0 = (\tilde{\pi}_0(\mathcal{A}), \tilde{\pi}_0(0), \tilde{\pi}_0(1))^T = (0,1,0)^T.
\end{equation}
Thus, the Markov chain $\{\Theta_k\}$ starts at $0$. 
As long as the control $U_k=2$, which means to continue, the states evolve according
to the transition probability matrix $P_2$. The values selected for elements of matrix $P_2$ ensure that the absorption time to the state $1$ is inevitable and is a geometrically distributed 
random variable with parameter $\rho$. 
Before the absorption, the distributions of the observations are i.p.i.d. with law $(f_1, \cdots, f_T)$. After the state is absorbed in $1$
the distributions of the observations changes to that of an i.p.i.d. process with law $(g_1, \cdots, g_T)$. 

\medskip
\item \textit{Cost}: The cost $C_k(\theta,u)$ associated with state $\Theta_k=\theta$ and control $U_k=u$ is defined for $k \geq 0$ as
\begin{equation}
\begin{split}
C_k(0,1) &= \lambda_k\\
C_k(1,2) &= d_k \\
C_k(\theta,u) &= 0, \; \text{otherwise}.
\end{split}
\end{equation}
Thus, $\lambda_k$ is the penalty on the false alarm at time $k$ and $d_k$ is the penalty on the delay at time $k$. We assume that
the penalty sequences are periodic with period $T$: for any $k \geq 0$, 
\begin{equation}
\begin{split}
\lambda_{k+T} &= \lambda_k \\
d_{k+T} &= d_k.
\end{split}
\end{equation}

\medskip
\item \textit{Policy}: Let 
$$I_k = (Y_1, \cdots, Y_k, U_1, \cdots, U_{k-1})$$ 
be the information at time $k$. Also define a policy 
$$
\tilde{\Phi} = (\tilde{\phi}_1, \tilde{\phi}_2, \cdots)
$$ 
to be a sequence of mappings such that $U_k = \tilde{\phi}_k(I_k), \; \forall k$.
\end{itemize}

\medskip
We want to find a control policy to optimize the long term cost
\begin{equation}\label{eq:modShirPOMDP}
\begin{split}
V(\tilde{\pi}_0) = \min_{\tilde{\Phi}} \Expect\left[\sum_{k=0}^\infty C_k(\Theta_k, U_k)\right].
\end{split}
\end{equation}
Using arguments similar to those used to obtain \eqref{eq:ShirPOMDP}, it can be shown 
that the probability sequence 
$$
p_n = \Prob(\Theta_n=1 | Y_1, \cdots, Y_n) = \Prob(\nu \leq n | Y_1, \cdots, Y_n)
$$
is a sufficient statistics also for the problem in \eqref{eq:modShirPOMDP}. Consequently, solving \eqref{eq:modShirPOMDP} is equivalent 
to solving the following MDP problem:
\begin{equation}\label{eq:ModShirPOMDPproblem}
\min_{\tau} \; \Expect\left[ \sum_{n=0}^{\tau-1} d_n p_n + \lambda_\tau (1-p_\tau)\right].
\end{equation}

\medskip
If $\lambda_k = \lambda_f$ and $d_k =1$, $\forall k$, then the problem in \eqref{eq:ModShirPOMDPproblem} reduces to the problem in 
\eqref{eq:ShirPOMDP}. 
If the period $T=1$ and the processes are i.i.d., then the problem in \eqref{eq:ModShirPOMDPproblem} can also be solved using 
classical MDP theory \cite{veer-bane-elsevierbook-2013, krishnamurthy2016partially, bane-fusion-2018, bert-dyn-prog-book-2017}. 
However, for $T > 1$, the observation process is i.p.i.d. and the classical theory cannot be applied. 
The optimal control theory developed in Section~\ref{sec:MDP} below for periodic MDP can and will be used to solve 
\eqref{eq:ModShirPOMDPproblem} and hence its special case \eqref{eq:ShirPOMDP}. 

\section{Optimal Control of Periodic MDPs}\label{sec:MDP}
In this section, we develop an optimal control theory for MDPs with periodic cost and transition structure. 
We have stochastic processes $\{X_k\}$, $\{U_k\}$, and $\{W_k\}$ taking values is spaces as follows:
\begin{equation}\label{eq:assumptionXUW}
\begin{split}
X_k & \in \mathbb{R}, \quad \forall k, \\
U_k & \in \mathbb{U} = \{u_1, u_2, \cdots, u_m\}, \quad \forall k, \\
W_k & \in \mathbb{R}, \quad \forall k.
\end{split}
\end{equation}
The process $\{X_k\}$ is an MDP generated according to the transition structure
\begin{equation}
X_k = \phi_{k-1}\left( X_{k-1}, U_{k-1}, W_{k-1}\right), \quad k \geq 1. 
\end{equation}
Here $\{U_k\}$ is the control process and $\{W_k\}$ is the disturbance process. We assume that given $X_k=x_k$ and $U_k=u_k$, the distribution
of the disturbance $W_k$ is independent of the past disturbances $W_0, \cdots, W_{k-1}$. We use $t_k(w_k|x_k, u_k)$ to denote this conditional distribution. 
Thus, the state and disturbance spaces are real-valued and 
the control spaces are finite. The results in the paper are, in fact, valid for more general spaces. 
To accommodate more general spaces, the proof techniques may have to be slightly modified \cite{bert-dyn-prog-book-2017, bertsekasshreve1978}.

The main assumption in our model is that
the transition functions $\{\phi_k\}$ and the conditional distributions $\{t_k(w_k|x_k, u_k)\}$ are periodic: there is a positive integer $T$ such that for $k \geq 0$,
\begin{equation}
\begin{split}
\phi_{k+T}\left( X_{k+T}, U_{k+T}, W_{k+T}\right) &= \phi_k \left( X_{k+T}, U_{k+T}, W_{k+T}\right),\\
t_{k+T}(w_{k+T}|x_{k+T}, u_{k+T}) &= t_{k}(w_{k+T}|x_{k+T}, u_{k+T}) .
\end{split}
\end{equation}
The objective is to choose the control process so as to minimize the cost
\begin{equation}\label{eq:longtermaddcost}
\Expect \left[ \sum_{k=0}^\infty \alpha^k c_k (X_k, U_k, W_k)\right],
\end{equation}
where $\alpha \in [0,1]$ is the discount factor which is allowed to be equal to $1$ with a view towards problems in optimal stopping. 
The cost functions $\{c_k\}$ are assumed to be non-negative and periodic with the same period $T$: for $k \geq 0$,
\begin{equation}
\begin{split}
c_k(x,u,w) &\geq 0, \quad \forall x,u,w, \\ 
c_{k+T}\left( X_{k+T}, U_{k+T}, W_{k+T}\right) &= c_k \left( X_{k+T}, U_{k+T}, W_{k+T}\right).
\end{split}
\end{equation}
The assumption of non-negativity of the cost functions (which is the same as the assumption $P$ in \cite{bert-dyn-prog-book-2017}) 
ensures that all infinite summations are well-defined by monotone convergence theorem \cite{will-book-probmart-1991}. 

In order to minimize this long-term additive cost \eqref{eq:longtermaddcost}, we search over Markov control policies of type
$$
\Pi = \left[ \mu_0, \mu_1, \cdots \right]
$$
such that
$$
U_k = \mu_k(X_k), \quad k=0,1, \cdots.
$$
As done in \cite{bert-dyn-prog-book-2017}, it can be argued that restricting the search over Markov policies is sufficient. For a policy $\Pi$, 
we define the cost-to-go function starting with the state $X_0=x_0$ as
\begin{equation}\label{eq:V_PI}
V_\Pi(x_0) = \Expect \left[ \sum_{k=0}^\infty \alpha^k c_k \left(X_k, \mu_k(X_k), W_k\right) \bigg| X_0=x_0\right],
\end{equation}
where the expectation is with respect to the disturbances. We are interested in solving the following problem: for $x_0 \in \mathbb{R}$,
\begin{equation}\label{eq:perMDPprob}
\begin{split}
V^*(x_0) &= \min_\Pi V_\Pi(x_0) \\
&= \min_\Pi \Expect \left[ \sum_{k=0}^\infty \alpha^k c_k \left(X_k, \mu_k(X_k), W_k\right)\bigg| X_0=x_0\right].
\end{split}
\end{equation}
In this section, we show that the optimal policy for the problem in \eqref{eq:perMDPprob} is periodic with period $T$, i.e., it is of the type
$$
\Pi^* = \left[ \mu_0^*, \cdots \mu_{T-1}^*, \mu_0^*, \cdots, \mu_{T-1}^*, \cdots \right].
$$
We also provide an explicit way to compute this optimal periodic policy.


For $\ell =0, 1, \cdots, T-1$, and $x \in \mathbb{R}$, define the operator
\begin{equation}\label{eq:operatormin}
\Psi^{(\ell)}(J)(x) = \min_{u\in \mathbb{U}} \; \Expect^{(\ell)} \left[ c_\ell(x, u, W) + \alpha J(\phi_\ell(x, u, W)) \right],
\end{equation}
where the expectation $\Expect^{(\ell)} $ is defined with respect to the conditional distribution $t_\ell(w | x,u)$. We also define 
the operator for a Markov map $\mu$ and $x \in \mathbb{R}$:
\begin{equation}\label{eq:operatormu}
\Psi^{(\ell)}_\mu(J)(x) = \Expect^{(\ell)} \left[ c_\ell(x, \mu(x), W) + \alpha J(\phi_\ell(x, \mu(x), W)) \right].
\end{equation}
Finally, define the $T-$fold operator
\begin{equation}\label{eq:prodofPsis}
\Psi =  \Psi^{(0)} \Psi^{(1)} \cdots \Psi^{(T-1)},
\end{equation}
which is the successive application of the $T$ operators defined in \eqref{eq:operatormin}. 
Our first result is the following. 

\medskip
\begin{theorem}\label{thm:Bellman}
The optimal cost function $V^*$ in \eqref{eq:perMDPprob} satisfies the following fixed-point equation: for any $x \in \mathbb{R}$,
\begin{equation}\label{eq:BellmanPeriod}
V^*(x) = (\Psi)(V^*)(x) = \Psi^{(0)} \Psi^{(1)} \cdots \Psi^{(T-1)}(V^*)(x).   
\end{equation}
\end{theorem}
\begin{IEEEproof}
The proof is provided in the appendix. 
\end{IEEEproof}
\medskip

Next, we show that if the optimal cost function is known, then the optimal policy can be obtained and shown to be periodic.
\begin{theorem}\label{thm:OptimalPolicyper}
The optimal policy is periodic. Specifically, let $V^*$ be the optimal cost function and let $[\mu^*_0, \cdots \mu^*_{T-1}]$ be such that 
for $\ell =0, 1, \cdots, T-1$, and $x \in \mathbb{R}$,
\begin{equation}\label{eq:muellopt}
\Psi^{(\ell)}_{\mu^*_\ell}\left(\Psi^{(\ell+1)} \cdots \Psi^{(T-1)}(V^*)\right)(x) = \Psi^{(\ell)} \cdots \Psi^{(T-1)}(V^*)(x).
\end{equation}
Then, the optimal policy is given by
\begin{equation}\label{eq:PeriodicOptPolicPistar}
\Pi^* = \left[ \mu_0^*, \cdots \mu_{T-1}^*, \mu_0^*, \cdots, \mu_{T-1}^*, \cdots \right].
\end{equation}
\end{theorem}
\begin{IEEEproof}
The proof is provided in the appendix. 
\end{IEEEproof}
\medskip
An optimal policy always exists in our case because we assume the control spaces to be finite. 
\medskip

The previous result is useful only when we have an algorithm to compute the optimal cost function $V^*$. This is facilitated by  
the theorem below. Define
\begin{equation}\label{eq:valiteJk}
V_k = \left[ \Psi  \right]^k (V_0),
\end{equation}
where we have used $\left[ \Psi  \right]^k(V_0)$ to denote the operator $\Psi$ in \eqref{eq:prodofPsis} applied $k$ times to the all-zero function $V_0$.
\begin{theorem}\label{thm:valueiteration}
The limit of sequence $\{V_k\}$ in \eqref{eq:valiteJk} exists:
\begin{equation}\label{eq:ValueIteration}
V_\infty = \lim_{k \to \infty} V_k =  \lim_{k \to \infty}\left[ \Psi  \right]^k (V_0).
\end{equation}
Furthermore, 
$$
V^* = V_\infty.
$$
\end{theorem}
\begin{IEEEproof}
The proof is provided in the appendix. 
\end{IEEEproof}

\medskip
Thus, the recipe for finding the periodic optimal policy is to start with the all-zero function and apply the value iteration \eqref{eq:valiteJk} to obtain the optimal 
cost function $V^* = V_\infty$. Finally, solve \eqref{eq:muellopt} to obtain the $T$ Markov maps  $[\mu^*_0, \cdots \mu^*_{T-1}]$ to construct the optimal policy $ \Pi^*$ in \eqref{eq:PeriodicOptPolicPistar}.  
 
\section{Detecting Changes in I.P.I.D. Processes}\label{sec:ApplnOST}
In this section, we solve the problem in \eqref{eq:ModShirPOMDPproblem}, and hence the problem in \eqref{eq:ShirPOMDP}, using 
Theorems~\ref{thm:Bellman}, \ref{thm:OptimalPolicyper}, and \ref{thm:valueiteration}. 
Specifically, we can state the following result. Let 
\begin{equation}\label{eq:optJipid}
J^*(p) =  \min_{\tau} \; \Expect\left[ \sum_{n=0}^{\tau-1} d_n p_n + \lambda_\tau (1-p_\tau)\right],\; p \in [0,1],
\end{equation}
where $p_0=p$.

\medskip
\begin{theorem}\label{thm:ipidOpt}
The optimal cost function in \eqref{eq:optJipid} satisfies
\begin{equation}\label{eq:Bellmanipid}
J^*(p) = \Psi^{(0)} \Psi^{(1)} \cdots \Psi^{(T-1)}(J^*)(p),    \; p \in [0,1],
\end{equation}
where for $\ell = 0, \cdots, T-1$,
\begin{equation}\label{eq:ipidTperrecursion}
\begin{split}
 \Psi^{(\ell)}(J)(p) = \min\left\{\lambda_\ell (1-p), \; p \; d_\ell  + A^{(\ell)}_J(p)\right\}.
\end{split}
\end{equation}
In the above equation, 
\begin{equation}\label{eq:ipidTperrecursionAJ}
\begin{split}
 A^{(\ell)}_J(p) = \int_x J\left( \bar{\phi}_{\ell}(p, x) \right) \left(\tilde{p} g_{\ell+1}(x) + (1-\tilde{p})f_{\ell+1}(x) \right)dx,
\end{split}
\end{equation}
where $\tilde{p} = p + (1-p) \rho$, and 
\begin{equation}\label{eq:phiforbelief}
\bar{\phi}_\ell(p, x) =  \frac{\tilde{p} \; g_{\ell+1}(x)}{\tilde{p} \; g_{\ell+1}(x)+ (1-\tilde{p}) f_{\ell+1}(x)}.
\end{equation} 
\end{theorem}
\medskip

In the above theorem, the densities are assumed to be with respect to the Lebesgue measure. The expressions can be modified to allow for counting measures (summations) or 
more general measures. 

If $T=1$ and $d_0=1$, then we are reduced to the classical formulation of Shiryaev. It is well known that the optimal policy for the classical case is a single-threshold policy 
in which the change is declared the first time the statistic or probability $p_n$ is above a pre-defined threshold. 
The threshold depends on the choice of the false alarm penalty $\lambda_0$. 

For $T > 1$, it is interesting to ask if the Shiryaev single-threshold stopping rule is still the optimal policy. In the subsections below, we will show that the optimal policy, in fact,  
utilizes multiple thresholds, where the number of distinct thresholds can be up to $T$. 
For reference below, we define 
the Shiryaev stopping rule $\tau_{ps}$ here: for $A\geq 0$, 
\begin{equation}\label{eq:PeriodicShiryaev}
\tau_{ps} = \inf \{n \geq 0: p_n > A\}.
\end{equation}
In the rest of the paper, we call this stopping rule or policy the periodic-Shiryaev stopping rule or algorithm to emphasize that the recursion for $p_n$ is periodic.

\subsection{Example: Change Detection in I.P.I.D. Processes For Different Values of $T$}
In this section, we show two examples in the i.p.i.d. setting where the optimal policy is not stationary. In fact, it is periodic with periodic thresholds. 

\begin{figure}
    \centering
    \begin{subfigure}[b]{0.23\textwidth}
        \includegraphics[width=\textwidth]{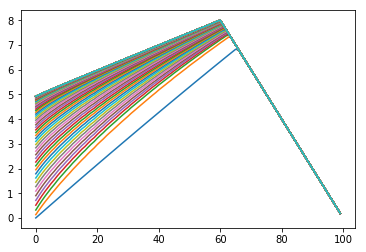}
        \caption{Cost curves at each stage of value iteration.}
        \label{fig:T2ValIte}
    \end{subfigure}
    ~ 
    \begin{subfigure}[b]{0.23\textwidth}
        \includegraphics[width=\textwidth]{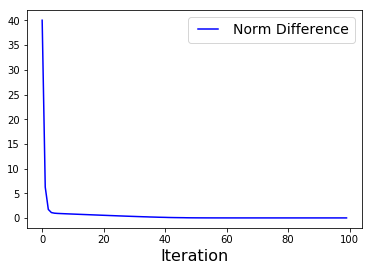}
        \caption{$L_2$ distance between successive cost functions in value iteration.}
        \label{fig:T2Norm}
    \end{subfigure}
    ~ 
    \begin{subfigure}[b]{0.23\textwidth}
        \includegraphics[width=\textwidth]{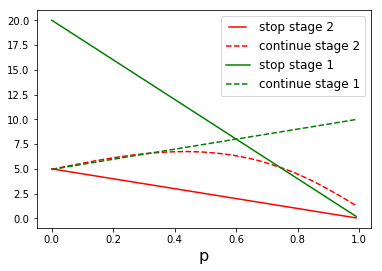}
        \caption{Optimal cost curves for the two stages.}
        \label{fig:T2Costs}
    \end{subfigure}
    ~ \begin{subfigure}[b]{0.23\textwidth}
        \includegraphics[width=\textwidth]{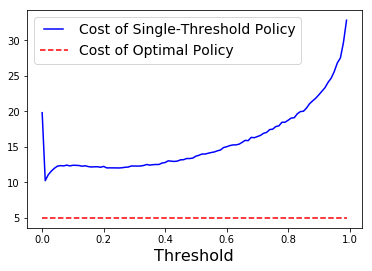}
        \caption{Performance of the single-threshold policy \eqref{eq:PeriodicShiryaev} as a function of the threshold used. }
        \label{fig:T2Comp}
    \end{subfigure}
    \caption{Cost functions and results for period $T=2$ and parameters in \eqref{eq:subsecT2_1} and \eqref{eq:subsecT2_2}. 
    The optimal cost function is obtained using value iteration (Theorem~\ref{thm:valueiteration}). Fig. (a) contains the plots of the cost function at each stage 
    of the iteration. Fig. (b) contains the plot of the norm distance between successive cost functions in the value iteration. Fig. (c) contains the optimal cost 
    curves for the two stages of this problem. The figures show that there are two thresholds $A_1$ and $A_2$. 
For stopping, $p_n$ is compared with the threshold $A_1$ during the odd times and compared with the threshold $A_2$ during the even times. The optimal cost achievable is $J^*(0)=5.0$ with thresholds $A_1=0.6$ and $A_2=0$. Fig. (d) contains the costs achievable by the single-threshold policy \eqref{eq:PeriodicShiryaev}. }\label{fig:T2Figures}
\end{figure}

In Fig.~\ref{fig:T2Figures}, we report results for $T=2$. Specifically, 
consider a change detection problem where the period of the i.p.i.d. processes is $T=2$ and the pre- and post-change i.p.i.d. densities are Gaussian:
\begin{equation}\label{eq:subsecT2_1}
\begin{split}
f_1 &= f_2 = \mathcal{N}(0,1),\\
g_1 &= \mathcal{N}(2,1),\\
g_2 &= \mathcal{N}(1,1).
\end{split}
\end{equation}
The parameters for change point, false alarm and delay are as follows:
\begin{equation}\label{eq:subsecT2_2}
\begin{split}
\lambda_0 &= 20, \quad \lambda_1 = 5\\
d_0 &=10, \quad d_1 =1\\
\rho &=0.01.
\end{split}
\end{equation}
The optimal cost function was obtained using value iteration \eqref{eq:ValueIteration} with the operators $\{\Psi^{(\ell)}\}$ as defined in \eqref{eq:ipidTperrecursion}. 
The cost functions at each iteration 
$$
J_k = \left[ \Psi  \right]^k (J_0), \quad \text{for} \; k \geq 1,
$$ 
with $J_0$ being the all-zero function on $[0,1]$, are plotted in Fig.~\ref{fig:T2ValIte}. We used a $100$ point resolution or discretization of the interval $[0,1]$ in the value iteration. 
In Fig~\ref{fig:T2Norm}, we have plotted the norm distance $\|J_k - J_{k+1}\|_2$ as a function 
of the iteration index $k$. In Fig.~\ref{fig:T2Costs}, we have plotted the stopping cost and the continue cost appearing in \eqref{eq:ipidTperrecursion} for $\ell =0,1$, 
where we use stages to refer to the $T$ distinct time slots. 
It can be inferred from Fig.~\ref{fig:T2Costs} that the optimal policy has alternating thresholds: $A_1=0.6$ and $A_2=0$. 
For stopping, $p_n$ is compared with the threshold $A_1$ during the odd time slots and compared with the threshold $A_2$ during the even time slots. The optimal cost achieved
by this alternating threshold policy is $J^*(0)=5.0$, as can be seen in Fig.~\ref{fig:T2ValIte}. In Fig.~\ref{fig:T2Comp}, we have plotted the total cost achieved by the 
periodic-Shiryaev algorithm \eqref{eq:PeriodicShiryaev} for different values of constant threshold $A$. These costs were obtained through Monte-Carlo simulations using $10,000$ sample paths. The best achievable cost of the single-threshold periodic-Shiryaev algorithm is $10.2$ establishing that the optimal cost of $5.0$ cannot be achieved using a single-threshold 
policy. 

\begin{figure}
    \centering
    \begin{subfigure}[b]{0.23\textwidth}
        \includegraphics[width=\textwidth]{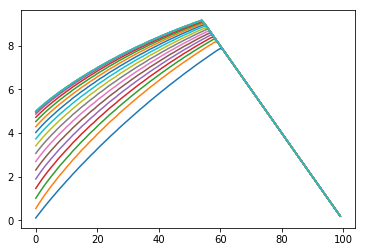}
        \caption{Cost curves at each stage of value iteration.}
        \label{fig:T4ValIte}
    \end{subfigure}
    ~ 
    \begin{subfigure}[b]{0.23\textwidth}
        \includegraphics[width=\textwidth]{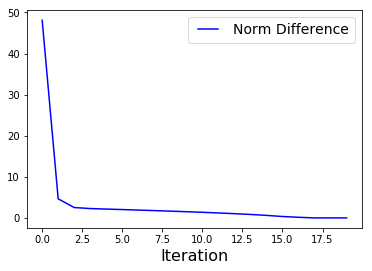}
        \caption{$L_2$ distance between successive cost functions in value iteration.}
        \label{fig:T4Norm}
    \end{subfigure}
    ~ 
    \begin{subfigure}[b]{0.23\textwidth}
        \includegraphics[width=\textwidth]{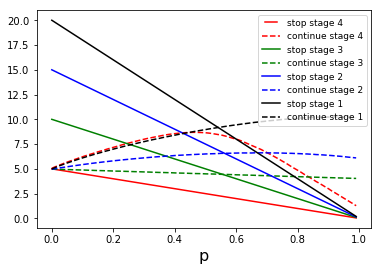}
        \caption{Optimal cost curves for the four stages.}
        \label{fig:T4Costs}
    \end{subfigure}
    ~ \begin{subfigure}[b]{0.23\textwidth}
        \includegraphics[width=\textwidth]{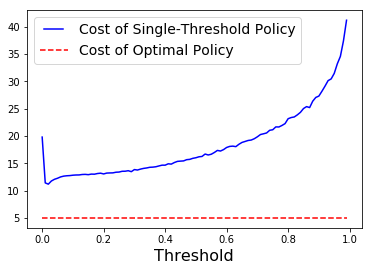}
        \caption{Performance of the single-threshold policy \eqref{eq:PeriodicShiryaev} as a function of the threshold used. }
        \label{fig:T4Comp}
    \end{subfigure}
    \caption{Cost functions and results for period $T=4$ and parameters in \eqref{eq:subsecT4_1} and \eqref{eq:subsecT4_2}. 
    Fig. (c) contains the optimal cost curves for the four stages of this problem. The figures show that there are four thresholds. 
The optimal cost achievable is $J^*(0)=5.0$. Fig. (d) contains the costs achievable by the single-threshold policy \eqref{eq:PeriodicShiryaev}.}\label{fig:T4figures}
\end{figure}

Next, we consider a change detection problem where the period of the i.p.i.d. processes is $T=4$ and the pre- and post-change i.p.i.d. densities are given by
\begin{equation}\label{eq:subsecT4_1}
\begin{split}
f_1 &= f_2 = f_3 = f_4 = \mathcal{N}(0,1),\\
g_1 &= \mathcal{N}(2,1), \quad g_2 = \mathcal{N}(1.5,1),\\
g_3 &= \mathcal{N}(1,1), \quad g_4 = \mathcal{N}(0.5,1).
\end{split}
\end{equation}
The parameters for change point, false alarm and delay are as follows:
\begin{equation}\label{eq:subsecT4_2}
\begin{split}
\lambda_0 &= 20, \quad \lambda_1 = 15, \quad \lambda_2=10, \quad \lambda_3=5,\\
d_0 &=10, \quad d_1 =10, \quad d_2=6, \quad d_3=1,\\
\rho &=0.01.
\end{split}
\end{equation}
Results for this case are reported in Fig.~\ref{fig:T4figures}. Similar to results for $T=2$ in Fig.~\ref{fig:T2Figures}, we see here also that the optimal cost is again $5.0$ (Fig.~\ref{fig:T4ValIte}). The Fig.~\ref{fig:T4Costs} also shows that there are four thresholds in this case, one for each of the four stages in a cycle or period of length $T=4$. Again, the best cost achievable by the periodic-Shiryaev algorithm equals $11.3$ and is strictly larger than the cost of the optimal policy. 

\medskip
\subsection{Change Detection in I.I.D. Data}
In this section, we show examples where the periodic-Shiryaev algorithm is not strictly optimal even for i.i.d. data, as long as $T>1$ and we use the modified Shiryaev 
formulation discussed in Section~\ref{sec:ModShir}. Specifically, we consider the change point problem with parameters
\begin{equation}\label{eq:subseciid_1}
\begin{split}
f_1 &= f_2 = \mathcal{N}(0,1),\\
g_1 &= g_2 = \mathcal{N}(\theta,1).
\end{split}
\end{equation}
and
\begin{equation}\label{eq:subseciid_2}
\begin{split}
\lambda_0 &= 20, \quad \lambda_1 = 5\\
d_0 &=10, \quad d_1 =1\\
\rho &=0.01.
\end{split}
\end{equation}
In Table~\ref{tab:iid}, we have reported the comparison between the optimal policy and the periodic-Shiryaev algorithm for the above parameters for different choices of 
the post-change parameter $\theta$. All points are obtained using Monte-Carlo simulations with $10,000$ sample paths. 
\begin{table}
\centering
\small
    \begin{tabular}{ | l | l | l | p{5cm} |}
    \hline
     & Cost of & Cost of\\ 
    $\theta$ & Periodic-Shiryaev & Optimal Policy \\ \hline\hline
    $0.5$ & $11.1$ & $5.0$\\ \hline
    $1.0$ & $12.0$ & $5.0$\\ \hline
    $2.0$ & $9.4$ & $5.0$\\ \hline     
    \end{tabular}
    \caption{Performance comparison for i.i.d. data.}
    \label{tab:iid}
\end{table}

\medskip

\subsection{Performance Comparison For Different Mean Choices}
In this section, we report comparison between the performance of the optimal policy and the periodic-Shiryaev algorithm for different choices of mean parameters for $T=2$:
\begin{equation}\label{eq:subsecdiffpara_1}
\begin{split}
f_1 &= f_2 = \mathcal{N}(0,1),\\
g_1 &= \mathcal{N}(\theta_1,1),\\
g_2 &= \mathcal{N}(\theta_2,1).
\end{split}
\end{equation}
The parameters for change point, false alarm and delay are as follows. 
\begin{equation}\label{eq:subsecdiffpara_2}
\begin{split}
\lambda_0 &= 20, \quad \lambda_1 = 5\\
d_0 &=10, \quad d_1 =1\\
\rho &=0.01.
\end{split}
\end{equation}
The results are collected in Table~\ref{tab:diffpara}. The values in the table suggests that the superiority of the optimal policy over the periodic-Shiryaev 
is maintained for different values of the mean parameters $\theta_1$ and $\theta_2$.  
\begin{table}
\centering
\small
    \begin{tabular}{|l | l | l | l | p{5cm} |}
    \hline
     && Cost of & Cost of\\ 
    $\theta_1$ &$\theta_2$& Single-Threshold Policy & Optimal Policy \\ \hline\hline
    $2.0$&$0.0$ & $7.2$ & $5.0$\\ \hline
       $2.0$&$0.5$ & $8.8$ & $5.0$\\ \hline
    $3.0$&$0.5$ & $6.6$ & $5.0$\\ \hline
    $3.0$&$1.0$ & $7.2$ & $5.0$\\ \hline
    $1.0$&$0.1$ & $9.5$ & $5.0$\\ \hline
    $0.5$&$0.0$ & $8.1$ & $5.0$\\ \hline
    \end{tabular}
    \caption{Performance comparison for different mean parameter choices.}
    \label{tab:diffpara}
\end{table}

\subsection{Performance Comparison For Different Choices of Delay and False Alarm Penalties}
In the previous sections, we have shown examples where the periodic-Shiryaev is strictly sub-optimal. In this section, we show
that the performance gap depends on the choice of the delay and false alarm penalties $\lambda_0, \lambda_1, d_0, d_1$. 
In Table~\ref{tab:diffdelaypfapara}, we show the performance comparison for 
\begin{equation}\label{eq:subsecdiffaddpfa_1}
\begin{split}
f_1 &= f_2 = \mathcal{N}(0,1),\\
g_1 &= \mathcal{N}(2,1),\\
g_2 &= \mathcal{N}(1,1).
\end{split}
\end{equation}
The performance gap reduces if the false alarm penalties are kept different but the delay penalties are set to the same value. The performance gap, in fact, vanishes 
as the false alarm and delay penalties are equal and the problem reduced to the classical Shiryaev case. In the next section, we provide a theoretical basis for this observation 
by showing that the periodic-Shiryaev algorithm is, in fact, asymptotically optimal for the classical Shiryaev formulation, as the false alarm rate goes to zero. 
While we could not find an example where the two algorithms have different performance for the classical Shiryaev formulation, we conjecture that such an example exists and necessarily involve high values of probability of false alarm. 
\begin{table}
\centering
\small
    \begin{tabular}{ | l | l | l | l | l | l | l | l |}
    \hline
                          &                       &            &            &                   &                   &            Cost of               &        Cost of\\ 
                          &                       &            &            &                   &                   & Single-Threshold          & Optimal  \\ 
    $\lambda_0$ & $\lambda_1$ & $d_0$ & $d_1$ & $\theta_1$ & $\theta_2$ &             Policy                &  Policy \\ \hline\hline
                 $20$ &      $5$           &  $10$   &  $1$          &  $2$                 &   $1$       & $10.2$                           & $5.0$\\ \hline
                 $20$ &      $5$           &  $1$   &  $1$          &  $2$                 &   $1$       & $4.6$                           & $3.7$\\ \hline
                 $5$ &      $5$           &  $1$   &  $1$          &  $2$                 &   $1$       & $3.2$                           & $3.2$\\ \hline
    \end{tabular}
    \caption{Performance comparison for different choices of delay and false alarm penalties.}
    \label{tab:diffdelaypfapara}
\end{table}

\section{Asymptotic Optimality of Single-Threshold Policies}\label{sec:Asymptotic}
In this section, we show that the periodic-Shiryaev algorithm is asymptotically optimal for the classical Shiryaev formulation 
\eqref{eq:ShiryProb} as the probability of false alarm goes to zero. 

For easy reference, we recall the definition of the periodic-Shiryaev algorithm here. Define
\begin{equation}\label{eq:ShirPeriodic}
p_n = \Prob^\pi(\nu \leq n | Y_1, \cdots, Y_n)
\end{equation}
and stop the first time this probability is above a threshold, i.e., use the stopping rule
\begin{equation}\label{eq:ShirStop}
\tau_{ps} = \min \{ n: p_n > A\}.
\end{equation}
While the statistic $p_n$ is always well-defined in a Bayesian setting, recall that 
in general for a non-i.i.d. model, the Shiryaev statistic cannot be computed recursively using a finite amount of memory \cite{tart-veer-siamtpa-2005, tart-niki-bass-2014}. 
Another convenient way to compute the statistic $p_n$ is to compute its transformation $R_n$ defined as
\begin{equation}\label{eq:ShirSR}
R_n = \frac{p_n}{1-p_n}. 
\end{equation}
The statistic $R_n$ can also be computed recursively. 

\medskip
\begin{lemma}\label{lem:recursion}
In the i.p.i.d. setting, the statistic $R_n$ \eqref{eq:ShirSR} can be computed recursively as 
\begin{equation}\label{eq:ShirSR_anyprior}
R_{n} = \left( R_{n-1} \frac{\Prob(\nu \geq n)}{\Prob(\nu > n)} + \frac{\pi_n}{\Prob(\nu > n)} \right) \frac{g_n(Y_n)}{f_n(Y_n)},
\end{equation}
with $R_0=0$.
Further, if the prior $\pi$ is $\text{Geom}(\rho)$ then the above recursion simplifies to 
\begin{equation}\label{eq:ShirSR_geomprior}
R_{n} = \left( \frac{R_{n-1} + \rho}{1-\rho}\right) \frac{g_n(Y_n)}{f_n(Y_n)}. 
\end{equation}
\end{lemma}
\begin{IEEEproof}
The proof is provided in the appendix.
\end{IEEEproof}
\medskip

\subsection{Universal Performance Bounds For Change Detection in I.P.I.D. Processes}
In this section, we obtain a universal lower bound on the performance of any stopping rule for detecting changes in an i.p.i.d. process. 

Let there exist $d\geq 0$ such that  
\begin{equation}\label{eq:priortail}
\lim_{n \to \infty} \frac{\log \Prob(\nu > n)}{n} = -d.
\end{equation}
If $\pi = \text{Geom}(\rho)$, then
$$
\frac{\log \Prob(\nu > n)}{n}  = \frac{\log (1-\rho)^n}{n} = \frac{n \log (1-\rho)}{n} = \log(1-\rho).  
$$
Thus, $d = |\log(1-\rho)|$.

Further, let 
\begin{equation}\label{eq:KLnumber}
I = \frac{1}{T}\sum_{i=1}^T D(g_i \; \| \; f_i),
\end{equation}
where $D(g_i \; \| \; f_i)$ is the Kullback-Leibler divergence between the densities $g_i$ and $f_i$. We assume 
that
$$
D(g_i \; \| \; f_i) < \infty, \quad \forall i=1, 2, \cdots, T,
$$
and
$$
0 < D(g_i \; \| \; f_i), \quad \text{ for some } i=1, 2, \cdots, T.
$$

\begin{theorem}\label{thm:LB}
Let the information number $I$ be as defined in \eqref{eq:KLnumber} and satisfy $0 < I < \infty$. Also, let $d$ be as in \eqref{eq:priortail}. 
Then, for any stopping time $\tau \in \mathbf{C}_\alpha$, i.e., for any $\tau$ satisfying the false alarm constraint $\Prob^\pi(\tau < \nu) \leq \alpha$, we 
have
\begin{equation}
\Expect^\pi\left[ \tau - \nu | \tau \geq \nu \right] \geq \frac{|\log \alpha| }{I + d}(1+o(1)), \quad \text{ as } \alpha \to 0.
\end{equation}
Here $o(1) \to 0$ as $\alpha \to 0$.
\end{theorem}
\begin{IEEEproof}
The proof is provided in the appendix.
\end{IEEEproof}

\subsection{Optimality of Periodic-Shiryaev Algorithm}
We now show that the periodic-Shiryaev algorithm \eqref{eq:ShirStop} is asymptotically optimal for problem \eqref{eq:ShiryProb} as the false 
alarm constraint $\alpha \to 0$. We will establish the optimality by showing that the periodic-Shiryaev algorithm achieves the lower bound specified
in Theorem~\ref{thm:LB}. 

Let 
$$
Z_i = \log \frac{g_i(Y_i)}{f_i(Y_i)},
$$
and define for $\epsilon > 0$ 
\begin{equation}\label{eq:LastTimeVar}
L_\epsilon^{(k)} = \sup \left\{n \geq 1: \frac{1}{n} \bigg| \sum_{i=k}^{k+n-1} Z_i - I \bigg| > \epsilon\right\}.
\end{equation}
We assume for simplicity that $\{f_i\}$ and $\{g_i\}$ are densities with respect to Lebesgue measure on the real line. The results below 
can be easily extended to densities with respect to more general measures (including the counting measure). 

\medskip
\begin{theorem}\label{thm:UB}
Let
\begin{equation}\label{eq:finitevarLLR}
\int_{-\infty}^\infty \left(\log \frac{g_i(y)}{f_i(y)}\right)^2 g_i(y)\;  dy < \infty, \quad \text{ for } i=1,2, \cdots, T.
\end{equation}
Then, we have
\begin{equation}\label{eq:FiniteLepsilon}
\begin{split}
\Expect_k \left[ L_\epsilon^{(k)} \right] &< \infty, \quad \forall \epsilon > 0, \; \forall k \geq 1, \\
\sum_{k=1}^\infty \pi_k \Expect_k \left[ L_\epsilon^{(k)} \right]  &< \infty, \quad \forall \epsilon > 0.
\end{split}
\end{equation}
The implication of \eqref{eq:FiniteLepsilon} is that with $A=1-\alpha$ in \eqref{eq:ShirStop}, we have
\begin{equation}\label{eq:ShirPerf}
\Expect^\pi\left[ \tau_{ps} - \nu | \tau_{ps} \geq \nu \right] \leq \frac{|\log \alpha| }{I + d}(1+o(1)), \quad \text{ as } \alpha \to 0.
\end{equation}
\end{theorem}
\begin{IEEEproof}
The proof is provided in the appendix.
\end{IEEEproof}

\medskip
Thus, the periodic-Shiryaev algorithm achieves the lower bound and is asymptotically optimal. The arguments provided 
in the proofs of the theorems above can be extended to also establish asymptotic optimality with respect to higher order moments of the detection delay. 

\subsection{Numerical Results}
\begin{figure}
\centering
\includegraphics[scale=0.5]{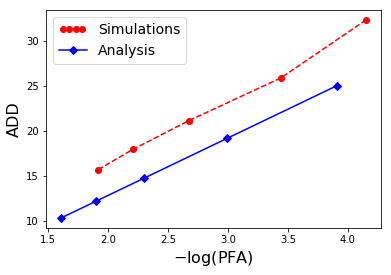}
\caption{Delay-False alarm trade-off curves for the periodic-Shiryaev algorithm. }
\label{fig:ADDPFA}
\end{figure}

\begin{figure}
\centering
\includegraphics[scale=0.5]{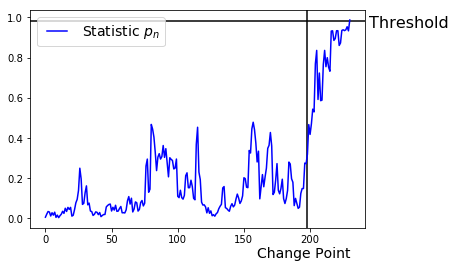}
\caption{Evolution of the periodic-Shiryaev statistic. }
\label{fig:PnEvolution}
\end{figure}

In Fig.~\ref{fig:ADDPFA}, we have plotted the average detection delay (ADD) $\Expect[(\tau_{ps}-\nu)^+]$ as a function of the magnitude of the logarithm of the probability of false alarm 
(PFA) for the following set of parameters: 
\begin{equation}\label{eq:addpfaspecs}
\begin{split}
f_1 &= f_2 = \mathcal{N}(0,1),\\
g_1 &= \mathcal{N}(0.75,1),\\
g_2 &= \mathcal{N}(0.25,1)\\
\rho &=0.01.
\end{split}
\end{equation}
The values for simulations were obtained using $5000$ sample paths. The values for the analysis curve in the figure were obtained by setting 
the probability of false alarm using the threshold $A=1-\alpha$ and using the delay expression $\frac{|\log \alpha| }{I + |\log(1-\rho)|}$. As can be observed from the figure, 
the analytical expression provides an accurate estimate of the delay. 
In Fig.~\ref{fig:PnEvolution}, we have plotted a typical sample path of the periodic-Shiryaev algorithm for the same set of parameters used in Fig.~\ref{fig:ADDPFA}. 

\section{Conclusion}
We established the optimality of periodic policies for optimal control of MDPs in which the cost structure 
and the transition probabilities are periodic, all with the same period. We then applied this result to solve an optimal stopping problem 
using the framework of partially observable MDPs. The optimal stopping problem we studied is the problem of detecting changes in 
i.p.i.d. processes. The exact optimality theory suggests that the optimal policy has multiple thresholds (alternating, in case the period $T=2$). 
This structural behavior, or its effect, 
is absent in the low false alarm regime, where we showed that using a single, fixed, threshold is asymptotically optimal, as the probability 
of false alarm goes to zero. A Bayesian analysis often provides important insights into a problem. The insight we obtain from this paper is 
that when analyzing a non-Bayesian or minimax version of the problem studied in this paper, one can conjecture that a single-threshold policy like cumulative sum is 
not strictly optimal for all values of the false alarm rate \cite{mous-astat-1986}. 

%
%

\appendix

\begin{IEEEproof}[Proof of Lemma~\ref{lem:ShirRecPn}]
For $k \leq n$ we have
\begin{equation*}
\begin{split}
& \Prob (\nu =k | Y_1, \cdots, Y_n) \\
&= \frac{\pi_k \prod_{i=1}^{k-1} f_i(Y_i) \prod_{i=k}^{n} g_i(Y_i)}{ \sum_{j=1}^n \pi_j \prod_{i=1}^{j-1} f_i(Y_i) \prod_{i=j}^{n} g_i(Y_i) + \Gamma_n \prod_{i=1}^{n} f_i(Y_i) },
\end{split}
\end{equation*}
where $\Gamma_n = \Prob(\nu > n)$. This can be formally proved using the rigorous definition of conditional expectations using sub-sigma-algebras \cite{will-book-probmart-1991}. 
This implies
\begin{equation*}
\begin{split}
&p_n=\Prob (\nu \leq n | Y_1, \cdots, Y_n) \\
&=  \frac{ \sum_{k=1}^n \pi_k \prod_{i=1}^{k-1} f_i(Y_i) \prod_{i=k}^{n} g_i(Y_i)}{ \sum_{k=1}^n \pi_k \prod_{i=1}^{k-1} f_i(Y_i) \prod_{i=k}^{n} g_i(Y_i) + \Gamma_n \prod_{i=1}^{n} f_i(Y_i) }.
\end{split}
\end{equation*}
Using this we can obtain an expression for $1-p_n$:
\begin{equation*}
\begin{split}
&1-p_n=\\
&=  \frac{\Gamma_n \prod_{i=1}^{n} f_i(Y_i) }{ \sum_{k=1}^n \pi_k \prod_{i=1}^{k-1} f_i(Y_i) \prod_{i=k}^{n} g_i(Y_i) + \Gamma_n \prod_{i=1}^{n} f_i(Y_i) }.
\end{split}
\end{equation*}
The numerator of $p_n$ can be written as
\begin{equation*}
\begin{split}
&\sum_{k=1}^n \pi_k \prod_{i=1}^{k-1} f_i(Y_i) \prod_{i=k}^{n} g_i(Y_i) \\
&=\sum_{k=1}^{n-1} \pi_k \prod_{i=1}^{k-1} f_i(Y_i) \prod_{i=k}^{n} g_i(Y_i) + \pi_n  \prod_{i=1}^{n-1} f_i(Y_i) g_n(Y_n) \\
&=\left(\sum_{k=1}^{n-1} \pi_k \prod_{i=1}^{k-1} f_i(Y_i) \prod_{i=k}^{n-1} g_i(Y_i) + \pi_n  \prod_{i=1}^{n-1} f_i(Y_i) \right)g_n(Y_n) 
\end{split}
\end{equation*}
Using this expansion and dividing the numerator and the denominator of $p_n$ by 
$\sum_{k=1}^{n-1} \pi_k \prod_{i=1}^{k-1} f_i(Y_i) \prod_{i=k}^{n-1} g_i(Y_i) + \Gamma_{n-1} \prod_{i=1}^{n-1} f_i(Y_i) $
we get
\begin{equation*}
\begin{split}
p_n= \frac{\left(p_{n-1} + (1-p_{n-1})\frac{\pi_n}{\Gamma_{n-1}}\right) \frac{g_n(Y_n)}{f_n(Y_n)}}{\left(p_{n-1} + (1-p_{n-1})\frac{\pi_n}{\Gamma_{n-1}}\right) + 
\frac{\Gamma_{n}}{\Gamma_{n-1}} (1-p_{n-1})}.
\end{split}
\end{equation*}
The lemma is proved if we now used the geometric prior. 

\end{IEEEproof}

\begin{IEEEproof}[Proof of Theorem~\ref{thm:Bellman}]
Recall from \eqref{eq:V_PI} that for any policy $\Pi=[\mu_0, \mu_1, \cdots]$, the cost-to-go function is given by
\begin{equation*}
V_\Pi(x_0) = \Expect \left[ \sum_{k=0}^\infty \alpha^k c_k \left(X_k, \mu_k(X_k), W_k\right) \bigg| X_0=x_0\right].
\end{equation*}
Rearranging terms we get
\begin{equation}
\begin{split}
&V_\Pi  (x_0) = \Expect \left[ \sum_{k=0}^\infty \alpha^k c_k \left(X_k, \mu_k(X_k), W_k\right) \bigg| X_0=x_0\right] \\
&= \Expect \left[ \sum_{k=0}^{T-1} \alpha^k c_k \left(X_k, \mu_k(X_k), W_k\right) \right. \\
&\quad \quad \quad \left. + \; \alpha^T \sum_{k=T}^\infty \alpha^{k-T} c_k \left(X_k, \mu_k(X_k), W_k\right) \bigg| X_0=x_0\right] \\
&=  \Expect \left[ \sum_{k=0}^{T-1} \alpha^k c_k \left(X_k, \mu_k(X_k), W_k\right) + \alpha^T \tilde{V}(X_T) \bigg| X_0=x_0\right],
\end{split}
\end{equation}
where for any $x \in \mathbb{R}$
$$
\tilde{V}(x) =  \Expect \left[ \sum_{k=T}^\infty \alpha^{k-T} c_k \left(X_k, \mu_k(X_k), W_k\right) \bigg| X_T=x\right].
$$
Using the fact that the transition structures, conditional distribution of the disturbances, and the cost structures are periodic with period $T$, we have
$$
\tilde{V}(x) \geq V^*(x), \quad \forall x \in \mathbb{R}.
$$
From this we get
\begin{equation}
\begin{split}
&V_\Pi (x_0) \\
&= \Expect \left[ \sum_{k=0}^{T-1} \alpha^k c_k \left(X_k, \mu_k(X_k), W_k\right) + \alpha^T \tilde{V}(X_T) \bigg| X_0=x_0\right] \\
& \geq \Expect \left[ \sum_{k=0}^{T-1} \alpha^k c_k \left(X_k, \mu_k(X_k), W_k\right) + \alpha^T V^*(X_T) \bigg| X_0=x_0\right].
\end{split}
\end{equation}

Further, 
\begin{equation}
\begin{split}
&V_\Pi  (x_0) \\
&\geq  \Expect \left[ \sum_{k=0}^{T-1} \alpha^k c_k \left(X_k, \mu_k(X_k), W_k\right) + \alpha^T V^*(X_T) \bigg| X_0=x_0\right]\\
&= \Expect \Bigg[ \sum_{k=0}^{T-2} \alpha^k c_k \left(X_k, \mu_k(X_k), W_k\right) \\
& \quad  \quad+ \; \alpha^{T-1} \bigg( c_{T-1} \left(X_{T-1}, \mu_{T-1}(X_{T-1}), W_{T-1} \right)  \\
& \quad  \quad \quad\quad\quad\quad\quad+ \alpha V^*(X_T)  \bigg) \bigg| X_0=x_0\Bigg]\\
&= \Expect \Bigg[ \sum_{k=0}^{T-2} \alpha^k c_k \left(X_k, \mu_k(X_k), W_k\right) \\
& \quad  \quad+ \; \alpha^{T-1} \Expect\bigg( c_{T-1} \left(X_{T-1}, \mu_{T-1}(X_{T-1}), W_{T-1} \right)  \\
& \quad  \quad \quad\quad\quad\quad\quad+ \alpha V^*(X_T) \bigg| X_{T-1} \bigg) \bigg| X_0=x_0\Bigg]\\
&= \Expect \left[ \sum_{k=0}^{T-2} \alpha^k c_k \left(X_k, \mu_k(X_k), W_k\right) \right. \\
& \quad \quad \quad \left. + \; \alpha^{T-1}\Psi_{\mu_{T-1}}^{(T-1)}\left(V^*\right)(X_{T-1}) \bigg| X_0=x_0\right]\\
&\geq \Expect \left[ \sum_{k=0}^{T-2} \alpha^k c_k \left(X_k, \mu_k(X_k), W_k\right) \right. \\
& \quad \quad \quad \left. + \; \alpha^{T-1}\Psi^{(T-1)}\left(V^*\right)(X_{T-1}) \bigg| X_0=x_0\right],
\end{split}
\end{equation}
where the operator $\Psi^{(T-1)}$ is defined in \eqref{eq:operatormin}. 
Continuing in this manner, we get
\begin{equation}
\begin{split}
V_\Pi  (x_0) &\geq \Expect \left[ c_0 \left(X_0, \mu_0(X_0), W_0\right) \right. \\
& \quad \quad \quad \left. + \; \alpha \Psi^{(1)} \Psi^{(2)}\cdots \Psi^{(T-1)}\left(V^*\right)(X_1) \bigg| X_0=x_0\right] \\
& = \Psi_{\mu_0}^{(0)} \Psi^{(1)} \Psi^{(2)}\cdots \Psi^{(T-1)}\left(V^*\right)(x_0)\\
& \geq \Psi^{(0)} \Psi^{(1)} \Psi^{(2)}\cdots \Psi^{(T-1)}\left(V^*\right)(x_0).
\end{split}
\end{equation}
Thus, for any $x \in \mathbb{R}$, we have
$$
V_\Pi  (x) \geq  \Psi^{(0)} \Psi^{(1)} \Psi^{(2)}\cdots \Psi^{(T-1)}\left(V^*\right)(x).
$$
Since the right hand side is not a function of the policy $\Pi$, we can take the minimum over all policies on the left to get
\begin{equation}\label{eq:VstartLB}
V^*(x) = \min_\Pi V_\Pi  (x) \; \geq \;  \Psi^{(0)} \Psi^{(1)} \Psi^{(2)}\cdots \Psi^{(T-1)}\left(V^*\right)(x).
\end{equation}

We now prove that the inequality above \eqref{eq:VstartLB} is also true in the other direction.  Let $[\mu^*_0, \cdots \mu^*_{T-1}]$ be such that $\forall x \in \mathbb{R}$
and for $\ell = 0, 1, \cdots, T-1$,
\begin{equation}
\Psi^{(\ell)}_{\mu^*_\ell}\left(\Psi^{(\ell+1)} \cdots \Psi^{(T-1)}(V^*)\right)(x) = \Psi^{(\ell)} \cdots \Psi^{(T-1)}(V^*)(x).
\end{equation}
Thus, $\mu^*_0, \cdots \mu^*_{T-1}$ are the optimal solutions to the minimization problem encountered in the definition of operators 
$\Psi^{(0)}, \cdots,  \Psi^{(T-1)}$, respectively. These optimal solutions always exist because the control space $\mathbb{U}$ is finite. 
Now consider the policy $\Pi^*$ given by
\begin{equation}
\Pi^* = \left[ \mu_0^*, \cdots \mu_{T-1}^*, \mu_0^*, \cdots, \mu_{T-1}^*, \cdots \right].
\end{equation}
By definition we have
$$
V^*(x) \leq V_{\Pi^*}(x), \quad \forall x \in \mathbb{R}.
$$
We now show that $V_{\Pi^*} \leq \Psi^{(0)} \Psi^{(1)} \Psi^{(2)}\cdots \Psi^{(T-1)}\left(V^*\right)$ thus proving the reverse inequality in \eqref{eq:VstartLB}.

Now, note that by the monotone convergence theorem
\begin{equation}
\begin{split}
&V_{\Pi^*}(x) = \Expect \left[ \sum_{k=0}^\infty \alpha^k c_k \left(X_k, \mu_k^*(X_k), W_k\right) \bigg| X_0=x\right]\\
&=\lim_{N \to \infty} \Expect \left[ \sum_{k=0}^N \alpha^k c_k \left(X_k, \mu_k^*(X_k), W_k\right) \bigg| X_0=x\right]
\end{split}
\end{equation}
Since the limit always exists, we can take the limit through a subsequence:
\begin{equation}
\begin{split}
&V_{\Pi^*}(x) \\
&=\lim_{m \to \infty, m \in \mathbb{N}} \Expect \left[ \sum_{k=0}^{mT-1} \alpha^k c_k \left(X_k, \mu_k^*(X_k), W_k\right) \bigg| X_0=x\right],
\end{split}
\end{equation}
where $\mathbb{N}$ is the set of positive integers. Now note that for any $m$
\begin{equation}
\begin{split}
\Expect \Bigg[ &\sum_{k=0}^{mT-1} \alpha^k c_k \left(X_k, \mu_k^*(X_k), W_k\right) \bigg| X_0=x\Bigg] \\
&\leq  \Expect \Bigg[ \sum_{k=0}^{mT-1} \alpha^k c_k \left(X_k, \mu_k^*(X_k), W_k\right) \\
&\quad \quad \quad \quad + \alpha^{mT} V^*(X_{mT})\bigg| X_0=x \Bigg] \\
&=\left[\Psi_{\mu_0^*}^{(0)} \Psi_{\mu_1^*}^{(1)} \Psi_{\mu_2^*}^{(2)}\cdots \Psi_{\mu_{T-1}^*}^{(T-1)}\right]^m(V^*)(x)\\
&=\left[\Psi_{\mu_0^*}^{(0)} \Psi_{\mu_1^*}^{(1)} \Psi_{\mu_2^*}^{(2)}\cdots \Psi_{\mu_{T-1}^*}^{(T-1)}\right]^{m-1}\\
& \quad \quad \quad \quad \Psi^{(0)} \Psi^{(1)} \Psi^{(2)}\cdots \Psi^{(T-1)}\left(V^*\right)(x)\\
&\leq \left[\Psi_{\mu_0^*}^{(0)} \Psi_{\mu_1^*}^{(1)} \Psi_{\mu_2^*}^{(2)}\cdots \Psi_{\mu_{T-1}^*}^{(T-1)}\right]^{m-1}(V^*)(x),
\end{split}
\end{equation}
where we have used the definition of $\Pi^*$ and \eqref{eq:VstartLB}. Also, we have used the notation $\left[\Psi_{\mu_0^*}^{(0)} \Psi_{\mu_1^*}^{(1)} \Psi_{\mu_2^*}^{(2)}\cdots \Psi_{\mu_{T-1}^*}^{(T-1)}\right]^m$ to denote that the composition of $T$ operators is applied $m$ times. 
Iterating in the above equation we get for any $m$
\begin{equation}
\begin{split}
\Expect \Bigg[ &\sum_{k=0}^{mT-1} \alpha^k c_k \left(X_k, \mu_k^*(X_k), W_k\right) \bigg| X_0=x\Bigg] \\
&\leq \left[\Psi_{\mu_0^*}^{(0)} \Psi_{\mu_1^*}^{(1)} \Psi_{\mu_2^*}^{(2)}\cdots \Psi_{\mu_{T-1}^*}^{(T-1)}\right]^{m-1}(V^*)(x) \\
&\leq \Psi^{(0)} \Psi^{(1)} \Psi^{(2)}\cdots \Psi^{(T-1)}\left(V^*\right)(x) \\
&\leq V^*(x).
\end{split}
\end{equation}
Since the upper bound is not a function of the limit index $m$, we get
\begin{equation}
\begin{split}
V^*(x) \leq &\;  V_{\Pi^*}(x) \\
=&\; \lim_{m \to \infty} \Expect \left[ \sum_{k=0}^{mT-1} \alpha^k c_k \left(X_k, \mu_k^*(X_k), W_k\right) \bigg| X_0=x\right]\\
\leq&  \; \Psi^{(0)} \Psi^{(1)} \Psi^{(2)}\cdots \Psi^{(T-1)}\left(V^*\right)(x) \\
\leq & \; V^*(x).
\end{split}
\end{equation}
Thus, all inequalities in the above equation are equality giving us
\begin{equation}\label{eq:VstartEQVpistart}
\begin{split}
V^*(x) =   V_{\Pi^*}(x) = \Psi^{(0)} \Psi^{(1)} \Psi^{(2)}\cdots \Psi^{(T-1)}\left(V^*\right)(x).
\end{split}
\end{equation}
This proves the theorem. 

\end{IEEEproof}
\medskip

\begin{IEEEproof}[Proof of Theorem~\ref{thm:OptimalPolicyper}]
The proof follows from \eqref{eq:VstartEQVpistart} in the proof of Theorem~\ref{thm:Bellman}. 
\end{IEEEproof}

\medskip

\begin{IEEEproof}[Proof of Theorem~\ref{thm:valueiteration}]
Recall that the operator $\Psi$ is defined as
$$
\Psi = \Psi^{(0)} \Psi^{(1)} \Psi^{(2)}\cdots \Psi^{(T-1)}.
$$
Also, we use the notation $\left[\Psi\right]^k$ to denote that the operator $\Psi$ is applied $k$ times. We are interested 
in the sequence of functions obtained by repeatedly applying the operator $\Psi$ to the all-zero function 
$V_0 \equiv 0$:
$$
V_k =  \left[ \Psi  \right]^k (V_0).
$$
First note that for any $\ell=0,1, \cdots, T-1$, and functions $J_1 \leq J_2$
\begin{equation}
\begin{split}
\Psi^{(\ell)}(J_1)(x) &= \min_{u\in \mathbb{U}} \; \Expect^{(\ell)} \left[ c_\ell(x, u, W) + \alpha J_1(\phi_\ell(x, u, W)) \right]\\
&\leq \min_{u\in \mathbb{U}} \; \Expect^{(\ell)} \left[ c_\ell(x, u, W) + \alpha J_2(\phi_\ell(x, u, W)) \right] \\
&=\Psi^{(\ell)}(J_2)(x), \quad \forall x \in \mathbb{R}.
\end{split}
\end{equation}
Since $V_0 \equiv 0$ and the cost functions are non-negative, we have $V_1 \geq V_0$. Using this and the monotonicity property established in the above display we obtain the following series of inequalities:
\begin{equation}
\begin{split}
V_1 &\geq V_0 \\
\Psi^{(T-1)}(V_1) &\geq \Psi^{(T-1)}(V_0) \\
\Psi^{(T-2)}\Psi^{(T-1)}(V_1) &\geq \Psi^{(T-2)}\Psi^{(T-1)}(V_0) \\
\vdots \\
\Psi^{(0)} \cdots \Psi^{(T-1)}(V_1) &\geq \Psi^{(0)} \cdots\Psi^{(T-1)}(V_0) \\
V_2 &\geq V_1.
\end{split}
\end{equation}
These equations establish that the sequence $V_k = \left[ \Psi  \right]^k (V_0)$ is a monotonically increasing sequence. Hence, the limit 
$$
V_\infty = \lim_{k\to \infty} V_k
$$
always exists. 

We next want to show that the limit $V_\infty$ satisfies the fixed-point equation $V_\infty = \Psi(V_\infty)$. Towards
this end, note that since $V_k \leq V_\infty$, we have
$$
V_{k+1} = \Psi(V_k) \leq \Psi(V_\infty).
$$
Taking the limit $k \to \infty$ we get
\begin{equation}\label{eq:VinfUB}
V_\infty \leq \Psi(V_\infty). 
\end{equation}
Let there exists $\tilde{x} \in \mathbb{R}$ such that
\begin{equation}\label{eq:VinfxtildstrctLess}
V_\infty(\tilde{x}) < \Psi(V_\infty)(\tilde{x}). 
\end{equation}
We now show that such an $\tilde{x}$ cannot exist. To show this, we evaluate $\left[ \Psi  \right]^{k+1} (V_0)$ 
at the value $\tilde{x}$: 
\begin{equation}\label{eq:VinffiniteUmin}
\begin{split}
 &\left[ \Psi  \right]^{k+1} (V_0)(\tilde{x}) = \Psi \left(\left[ \Psi  \right]^{k} (V_0)\right)(\tilde{x}) \\
 &= \Psi^{(0)} \Psi^{(1)} \Psi^{(2)}\cdots \Psi^{(T-1)} \left(\left[ \Psi  \right]^{k} (V_0)\right)(\tilde{x})\\
 &= \Psi^{(0)} \left(\Psi^{(1)} \Psi^{(2)}\cdots \Psi^{(T-1)} \left(\left[ \Psi  \right]^{k} (V_0)\right)\right)(\tilde{x})\\ 
 &=\min_{u\in \mathbb{U}} \; \Expect^{(0)} \Bigg[ c_0(\tilde{x}, u, W)  \\
 & \quad\quad\quad\quad \quad \quad + \alpha \Psi^{(1)} \cdots \Psi^{(T-1)} \left[ \Psi  \right]^{k} (V_0)(\phi_0(\tilde{x}, u, W)) \Bigg].
\end{split}
\end{equation}
Let $u_k(\tilde{x})$ be the control achieving the minimum in \eqref{eq:VinffiniteUmin}. Note that this optimal value depends 
on $\tilde{x}$ as well as the index $k$. Consider the sequence of these values $\{u_k(\tilde{x})\}$ as we vary $k$. Since the control space $\mathbb{U}$ is finite, there exists $\tilde{u} \in \mathbb{U}$ such that $u_k(\tilde{x}) = \tilde{u}$ infinite often, 
i.e., there exists a $\tilde{u} \in \mathbb{U}$ and an infinite set $\mathbb{N}_1 \subset \mathbb{N}$ such that 
$$
u_k(\tilde{x}) = \tilde{u}, \quad \text{ if } k \in \mathbb{N}_1.
$$
Thus, for $k \in \mathbb{N}_1$ we have
\begin{equation}\label{eq:VinffiniteUmin_2}
\begin{split}
 &\left[ \Psi  \right]^{k+1} (V_0)(\tilde{x}) \\
 &= \; \Expect^{(0)} \Bigg[ c_0(\tilde{x}, \tilde{u}, W)  \\
 & \quad\quad\quad\quad \quad \quad + \alpha \Psi^{(1)} \cdots \Psi^{(T-1)} \left[ \Psi  \right]^{k} (V_0)(\phi_0(\tilde{x}, \tilde{u}, W)) \Bigg].
\end{split}
\end{equation}
We wish to take $k \to \infty$ in the above equation by keeping $k$ in $\mathbb{N}_1$. To evaluate the limits in the above equation, we note that if ${J_k}$ is a sequence of non-negative increasing functions with monotone limit $J_k \uparrow J_\infty$, then for $\ell = 0, \cdots, T-1$,
\begin{equation}
\begin{split}
\Psi^{(\ell)}(J_k)(x) &= \; \min_{u\in \mathbb{U}} \; \Expect^{(\ell)} \left[ c_\ell(x, u, W) + \alpha J_k(\phi_\ell(x, u, W)) \right]\\
&\uparrow \; \min_{u\in \mathbb{U}} \; \Expect^{(\ell)} \left[ c_\ell(x, u, W) + \alpha J_\infty(\phi_\ell(x, u, W)) \right]\\
&=\Psi^{(\ell)}(J_\infty)(x),
\end{split}
\end{equation}
where we have used monotone convergence theorem and the fact that the control space $\mathbb{U}$ is finite. 
From this, we get the following set of limits:
\begin{equation}
\begin{split}
\left[ \Psi  \right]^{k} (V_0) \; &\uparrow \; V_\infty\\
 \Psi^{(T-1)} \left[ \Psi  \right]^{k} (V_0) \; &\uparrow \; \Psi^{(T-1)}(V_\infty)\\
  \Psi^{(T-2)} \Psi^{(T-1)} \left[ \Psi  \right]^{k} (V_0) \; &\uparrow \; \Psi^{(T-2)}\Psi^{(T-1)}(V_\infty)\\
  \vdots\\
  \Psi^{(1)} \cdots \Psi^{(T-1)} \left[ \Psi  \right]^{k} (V_0) \; &\uparrow \;   \Psi^{(1)} \cdots \Psi^{(T-1)} (V_\infty).
\end{split}
\end{equation}
Using the above limits and taking $k \to \infty$ in \eqref{eq:VinffiniteUmin_2} while $k \in \mathbb{N}_1$ we have from the monotone convergence theorem again 
\begin{equation}
\begin{split}
 &V_\infty(\tilde{x}) \\
 &= \; \Expect^{(0)} \Bigg[ c_0(\tilde{x}, \tilde{u}, W)  \\
 & \quad\quad\quad\quad \quad \quad + \alpha \Psi^{(1)} \cdots \Psi^{(T-1)} (V_\infty)(\phi_0(\tilde{x}, \tilde{u}, W)) \Bigg]\\
 & \geq \; \min_{u \in \mathbb{U}}\Expect^{(0)} \Bigg[ c_0(\tilde{x}, u, W)  \\
 & \quad\quad\quad\quad \quad \quad + \alpha \Psi^{(1)} \cdots \Psi^{(T-1)} (V_\infty)(\phi_0(\tilde{x}, u, W)) \Bigg]\\
 &=\;  \Psi^{(0)} \Psi^{(1)} \cdots \Psi^{(T-1)} (V_\infty)(\tilde{x})\\
 &=\; \Psi(V_\infty)(\tilde{x}).
\end{split}
\end{equation}
This contradicts \eqref{eq:VinfxtildstrctLess} and hence it must be true that
\begin{equation}\label{eq:VinfEQUPsiVinf}
V_\infty = \Psi(V_\infty) = \Psi^{(0)} \Psi^{(1)} \cdots \Psi^{(T-1)} (V_\infty).
\end{equation}
Thus, $V_\infty$ also satisfies the fixed-point equation that $V^*$ satisfies. 
If the solution to the fixed point equation is unique, then the theorem is proved. But, just as in the classical theory, which is obtained by setting $T=1$, the solution to the fixed point equation for $\alpha=1$ need not be unique. So, we now need an additional argument to show that $V_\infty=V^*$. 

We first show that $V^* \geq V_\infty$. This is true because $V^* \geq V_0$. Now, applying the $T$-fold operator $\Psi$ 
we get
$$
V^* = \Psi(V^*) \geq \Psi(V_0).
$$
Applying the operator $\Psi$ multiple times gives us
$$
V^* \geq \left[\Psi \right]^k(V_0) = V_k.
$$
Taking the limit $k \to \infty$ we get
$$
V^* \geq V_\infty. 
$$

To prove $V^* \leq V_\infty$, let $\mu^\infty_0, \cdots \mu^\infty_{T-1}$ be the optimal solutions to the minimization problem encountered in \eqref{eq:VinfEQUPsiVinf}. Again, these optimal solutions always exist because the control space $\mathbb{U}$ is finite. 
Now consider the policy $\Pi^\infty$ given by
\begin{equation}
\Pi^\infty = \left[ \mu_0^\infty, \cdots \mu_{T-1}^\infty, \mu_0^\infty, \cdots, \mu_{T-1}^\infty, \cdots \right].
\end{equation}
By definition we have
$$
V^*(x) \leq V_{\Pi^\infty}(x), \quad \forall x \in \mathbb{R}.
$$
We now show that $V_{\Pi^\infty} \leq V_\infty$ proving what we need: $V^* \leq V_\infty$. 
Towards this end, note that 
\begin{equation}
\begin{split}
&V_{\Pi^\infty}(x) \\
&=\lim_{m \to \infty} \Expect \left[ \sum_{k=0}^{mT-1} \alpha^k c_k \left(X_k, \mu_k^\infty(X_k), W_k\right) \bigg| X_0=x\right].
\end{split}
\end{equation}
Now note that for any $m$
\begin{equation}
\begin{split}
\Expect \Bigg[ &\sum_{k=0}^{mT-1} \alpha^k c_k \left(X_k, \mu_k^\infty(X_k), W_k\right) \bigg| X_0=x\Bigg] \\
&\leq  \Expect \Bigg[ \sum_{k=0}^{mT-1} \alpha^k c_k \left(X_k, \mu_k^\infty(X_k), W_k\right) \\
&\quad \quad \quad \quad + \alpha^{mT} V_\infty(X_{mT})\bigg| X_0=x \Bigg] \\
&=\left[\Psi_{\mu_0^\infty}^{(0)} \Psi_{\mu_1^\infty}^{(1)} \Psi_{\mu_2^\infty}^{(2)}\cdots \Psi_{\mu_{T-1}^\infty}^{(T-1)}\right]^m(V_\infty)(x)\\
&=\left[\Psi_{\mu_0^\infty}^{(0)} \Psi_{\mu_1^\infty}^{(1)} \Psi_{\mu_2^\infty}^{(2)}\cdots \Psi_{\mu_{T-1}^\infty}^{(T-1)}\right]^{m-1}\\
& \quad \quad \quad \quad \Psi^{(0)} \Psi^{(1)} \Psi^{(2)}\cdots \Psi^{(T-1)}\left(V_\infty\right)(x)\\
&= \left[\Psi_{\mu_0^\infty}^{(0)} \Psi_{\mu_1^\infty}^{(1)} \Psi_{\mu_2^\infty}^{(2)}\cdots \Psi_{\mu_{T-1}^\infty}^{(T-1)}\right]^{m-1}(V_\infty)(x),
\end{split}
\end{equation}
where we have used definition of $\Pi^\infty$ and the fact that $V_\infty$ satisfies $\Psi^{(0)} \Psi^{(1)} \Psi^{(2)}\cdots \Psi^{(T-1)}\left(V_\infty\right) = V_\infty$. 
Iterating in the above equation we get for any $m$
\begin{equation}
\begin{split}
\Expect \Bigg[ &\sum_{k=0}^{mT-1} \alpha^k c_k \left(X_k, \mu_k^\infty(X_k), W_k\right) \bigg| X_0=x\Bigg] \\
&\leq \left[\Psi_{\mu_0^\infty}^{(0)} \Psi_{\mu_1^\infty}^{(1)} \Psi_{\mu_2^\infty}^{(2)}\cdots \Psi_{\mu_{T-1}^\infty}^{(T-1)}\right]^{m-1}(V_\infty)(x) \\
&= \Psi^{(0)} \Psi^{(1)} \Psi^{(2)}\cdots \Psi^{(T-1)}\left(V_\infty\right)(x) \\
&= V_\infty(x).
\end{split}
\end{equation}
Since the upper bound is not a function of the limit index $m$, we get
\begin{equation}
\begin{split}
V^*(x) \leq &\;  V_{\Pi^\infty}(x) \\
=&\; \lim_{m \to \infty} \Expect \left[ \sum_{k=0}^{mT-1} \alpha^k c_k \left(X_k, \mu_k^\infty(X_k), W_k\right) \bigg| X_0=x\right]\\
\leq&  \; \Psi^{(0)} \Psi^{(1)} \Psi^{(2)}\cdots \Psi^{(T-1)}\left(V_\infty\right)(x) \\
= & \; V_\infty \\
\leq & \; V^*(x).
\end{split}
\end{equation}
Thus, all the inequalities in the above equation are equalities and the theorem is proved.

\end{IEEEproof}

\begin{IEEEproof}[Proof of Lemma~\ref{lem:recursion}]
Using the definition of $R_n = p_n/(1-p_n)$ we have 
\begin{equation}
\begin{split}
R_n &= \frac{ \sum_{k=1}^n \pi_k \prod_{i=1}^{k-1} f_i(Y_i) \prod_{i=k}^{n} g_i(Y_i)}{\Gamma_n \prod_{i=1}^{n} f_i(Y_i) }\\
&= \frac{1}{\Gamma_n}\sum_{k=1}^n \pi_k \prod_{i=k}^{n} \frac{ g_i(Y_i)}{f_i(Y_i) }.
\end{split}
\end{equation}
To get the recursion for $R_n$ we write
\begin{equation*}
\begin{split}
R_n &= \frac{1}{\Gamma_n}\sum_{k=1}^n \pi_k \prod_{i=k}^{n} \frac{ g_i(Y_i)}{f_i(Y_i) }\\
&= \frac{1}{\Gamma_n} \left(\sum_{k=1}^{n-1} \pi_k \prod_{i=k}^{n-1} \frac{ g_i(Y_i)}{f_i(Y_i) } + \pi_n\right) \frac{ g_n(Y_n)}{f_n(Y_n) }\\
&= \frac{1}{\Gamma_n} \left( \frac{\Gamma_{n-1}}{\Gamma_{n-1}} \sum_{k=1}^{n-1} \pi_k \prod_{i=k}^{n-1} \frac{ g_i(Y_i)}{f_i(Y_i) } + \pi_n\right) \frac{ g_n(Y_n)}{f_n(Y_n) }\\
&= \frac{1}{\Gamma_n} \left(\Gamma_{n-1} R_{n-1} + \pi_n \right)\frac{ g_n(Y_n)}{f_n(Y_n) }
\end{split}
\end{equation*}
proving \eqref{eq:ShirSR_anyprior}. Substituting for a geometric prior gives \eqref{eq:ShirSR_geomprior}. 
\end{IEEEproof}

\medskip
\medskip

\begin{IEEEproof}[Proof of Theorem~\ref{thm:LB}]
Let $Z_i = \log \frac{g_i(Y_i)}{f_i(Y_i)}$ be the log likelihood ratio at time $i$. We show that the sequence $\{Z_i\}$ satisfies the following statement:
\begin{equation}\label{eq:thm1_1}
\begin{split}
\lim_{n \to \infty}\Prob_\nu & \left( \frac{1}{n} \max_{t \leq n} \sum_{i=\nu}^{\nu+t} Z_i \geq I(1+\delta)\right) = 0, \; \forall \nu, \; \forall \delta > 0,
\end{split}
\end{equation}
where $I$ is as defined in \eqref{eq:KLnumber}. The lower bound then follows from Theorem 1 in \cite{tart-veer-siamtpa-2005}. 
Towards proving \eqref{eq:thm1_1}, note that as $n \to \infty$
\begin{equation}\label{eq:thm1_2}
\begin{split}
\frac{1}{n}\sum_{i=\nu}^{\nu+n} Z_i \to \frac{1}{T}\Expect_1\left[ \sum_{i=1}^T \log \frac{g_i(Y_i)}{f_i(Y_i)}\right] = I, \quad \text{a. s.} \; \Prob_\nu, \; \; \forall \nu \geq 1. 
\end{split}
\end{equation}
The above display is true because of the i.p.i.d. nature of the observation processes. This implies that as $n \to \infty$
\begin{equation}\label{eq:thm1_3}
\begin{split}
 \max_{t \leq n} \frac{1}{n}\sum_{i=\nu}^{\nu+t} Z_i \to I, \quad \text{a. s.} \; \Prob_\nu, \; \; \forall \nu \geq 1.
\end{split}
\end{equation}
To show this, note that
\begin{equation}\label{eq:thm1_4}
\begin{split}
 \max_{t \leq n} \frac{1}{n}\sum_{i=\nu}^{\nu+t} Z_i  = \max \left\{ \max_{t \leq n-1} \frac{1}{n}\sum_{i=\nu}^{\nu+t} Z_i, \; \;  \frac{1}{n}\sum_{i=\nu}^{\nu+n} Z_i\right\}.
\end{split}
\end{equation}
For a fixed $\epsilon > 0$, because of \eqref{eq:thm1_2} and the second term on the right, the LHS in \eqref{eq:thm1_3} is greater than $I(1-\epsilon)$ for $n$ large enough. Now, let the maximum on the LHS be achieved at a point $k_n$, 
then 
$$
 \max_{t \leq n} \frac{1}{n}\sum_{i=\nu}^{\nu+t} Z_i = \frac{1}{n}\sum_{i=\nu}^{\nu+k_n} Z_i  = \frac{k_n}{n} \frac{1}{k_n}\sum_{i=\nu}^{\nu+k_n} Z_i.
$$  
Now $k_n$ cannot be bounded because of the presence of $n$ in the denominator and the lower bound $I(1-\epsilon)$. This implies $k_n > i$, for any fixed $i$, and $k_n \to \infty$. Thus, $\frac{1}{k_n}\sum_{i=\nu}^{\nu+k_n} Z_i \to I$. Since $k_n/n \leq 1$, we have that the LHS in \eqref{eq:thm1_3} is less than $I(1+\epsilon)$, for $n$ large enough. This proves 
\eqref{eq:thm1_3}. 
To prove \eqref{eq:thm1_1}, note that due to the i.p.i.d. nature of the process
\begin{equation}\label{eq:thm1_5}
\begin{split}
\sup_{\nu \geq 1} \; \Prob_\nu & \left(  \frac{1}{n}\max_{t \leq n} \sum_{i=\nu}^{\nu+t} Z_i \geq I(1+\delta)\right) \\
&=\sup_{1 \leq \nu \leq T} \; \Prob_\nu  \left(  \frac{1}{n}\max_{t \leq n} \sum_{i=\nu}^{\nu+t} Z_i \geq I(1+\delta) \right) 
\end{split}
\end{equation}
The right hand side goes to zero because of \eqref{eq:thm1_3} (a.s. convergence implies convergence in probability) and because the maximum on the right hand side in \eqref{eq:thm1_5} is over only finitely many terms. This proves \eqref{eq:thm1_1} and the theorem. 
\end{IEEEproof}

\medskip
\medskip

\begin{IEEEproof}[Proof of Theorem~\ref{thm:UB}]
First consider the sequence
$$
\left\{\Expect_k \left[ L_\epsilon^{(k)} \right] \right\}_{k=1}^\infty.
$$
Due to the periodic nature of the process, the above sequence is also periodic with period $T$, and as a result, there are at most $T$ distinct values in the above sequence:
$$
\Expect_1 \left[ L_\epsilon^{(1)} \right], \cdots, \Expect_T \left[ L_\epsilon^{(T)} \right].
$$
Thus, if we show that these $T$ numbers are finite for any $\epsilon>0$, then we automatically have
$$
\sum_{k=1}^\infty \pi_k \Expect_k \left[ L_\epsilon^{(k)} \right]  < \infty, \quad \forall \epsilon > 0.
$$
Furthermore, since we do not make any explicit assumptions on the actual values taken by the densities $(f_1, \cdots, f_T)$
and $(g_1, \cdots, g_T)$, it is enough to show that 
$$
\Expect_1 \left[ L_\epsilon^{(1)} \right] < \infty, \quad \forall \epsilon > 0.
$$
This is because if we cyclically shift both the set of densities $(f_1, \cdots, f_T)$
and $(g_1, \cdots, g_T)$ by one, then $L_\epsilon^{(1)}$ with the new configuration of densities will be identically distributed  
as $L_\epsilon^{(2)}$ in the old configuration. 

Recall the definition of $L_\epsilon^{(1)} $:
$$
L_\epsilon^{(1)} = \sup \left\{n \geq 1: \frac{1}{n} \bigg| \sum_{i=1}^{n} Z_i - I \bigg| > \epsilon\right\}.
$$
To show that $L_\epsilon^{(1)}$ is integrable, note that for any $n$
$$
\left\{ \sup_{k \geq n } \bigg| \frac{1}{k} \sum_{i=1}^k Z_i - I\bigg| \leq \epsilon\right\} \subset \left\{L_\epsilon^{(1)} \leq n-1\right\}.
$$
Taking complements we get
$$
 \left\{L_\epsilon^{(1)} \geq n\right\} \subset \left\{ \sup_{k \geq n } \bigg| \frac{1}{k} \sum_{i=1}^k Z_i - I\bigg| > \epsilon\right\}.
$$
This implies
$$
\Prob_1\left(L_\epsilon^{(1)} \geq n \right) \leq \Prob_1\left(\sup_{k \geq n } \bigg| \frac{1}{k} \sum_{i=1}^k Z_i - I\bigg| > \epsilon \right).
$$
Summing over $n$ we get
\begin{equation}\label{thm2_1}
\begin{split}
\Expect_1 \left[ L_\epsilon^{(1)} \right] &= \sum_{n=1}^\infty \Prob_1\left(L_\epsilon^{(1)} \geq n \right) \\
&\leq \sum_{n=1}^\infty \Prob_1\left(\sup_{k \geq n } \bigg| \frac{1}{k} \sum_{i=1}^k Z_i - I\bigg| > \epsilon \right).
\end{split}
\end{equation}
To complete the theorem, we need to show that the rightmost term in \eqref{thm2_1} is finite. 

For $\ell =1, 2, \cdots, T$, define 
$$
Z_i^{(\ell)} = \begin{cases}
Z_i \quad \text{ if } i=mT+\ell \; \text{ for } m=0,1,2, \cdots\\
0 \quad \text{ otherwise, }
\end{cases}
$$
and
$$
I_\ell = D(g_\ell \; \| \; f_\ell).
$$
Note from \eqref{eq:KLnumber} that $I = \frac{I_1 + \cdots + I_T}{T}$.
Using these definitions, we can write $\frac{1}{k} \sum_{i=1}^k Z_i - I$ as
\begin{equation*}
\begin{split}
\frac{1}{k} \sum_{i=1}^k Z_i - I = \sum_{\ell =1}^T \left(\frac{1}{k} \sum_{i=1}^k Z_i^{(\ell)} - \frac{I_\ell}{T}\right).
\end{split}
\end{equation*}
From the above equation, we get for any $n$
\begin{equation*}
\begin{split}
\Prob_1 & \left(\sup_{k \geq n } \bigg| \frac{1}{k}  \sum_{i=1}^k Z_i - I\bigg| > \epsilon \right) \\
&\leq \sum_{\ell =1}^T  \Prob_1 \left(\sup_{k \geq n } \bigg| \frac{1}{k} \sum_{i=1}^k Z_i^{(\ell)} - \frac{I_\ell}{T}\bigg| > \frac{\epsilon}{T} \right).
\end{split}
\end{equation*}
Thus, to show the summability of the LHS in the above equation, we need to show the summability of each of the $T$ terms on the RHS. Again, due 
to the i.p.i.d. nature of the process, and because we have made no explicit assumptions about the densities $(f_1, \cdots, f_T)$
and $(g_1, \cdots, g_T)$, it is enough to establish the summability of the first term on the right.

Define for $\ell =1,2, \cdots, T$,
$$
I_i^{(\ell)} = \begin{cases}
I_\ell \quad \text{ if } i=mT+\ell \; \text{ for } m=0,1,2, \cdots\\
0 \quad \text{ otherwise.}
\end{cases}.
$$
Using this definition we write
$$
 \frac{1}{k} \sum_{i=1}^k Z_i^{(1)} - \frac{I_1}{I} =  \frac{1}{k} \sum_{i=1}^k (Z_i^{(1)} - I_i^{(1)} ) + \frac{1}{k} \sum_{i=1}^k I_i^{(1)} - \frac{I_1}{T}.
$$

Thus, with $\tilde{Z}_i^{(1)}  = Z_i^{(1)} - I_i^{(1)}$, we have 
\begin{equation}\label{eq:thm2_2}
\begin{split}
 \Prob_1 & \left(\sup_{k \geq n } \bigg| \frac{1}{k} \sum_{i=1}^k Z_i^{(1)} - \frac{I_1}{T}\bigg| > \frac{\epsilon}{T} \right) \\
 &\leq \Prob_1 \left(\sup_{k \geq n } \bigg| \frac{1}{k} \sum_{i=1}^k \tilde{Z}_i^{(1)}\bigg| > \frac{\epsilon}{2T} \right) \\
 & \quad + \Prob_1  \left(\sup_{k \geq n } \bigg| \frac{1}{k} \sum_{i=1}^k I_i^{(1)} - \frac{I_1}{T}\bigg| > \frac{\epsilon}{2T} \right).
\end{split}
\end{equation}
Now, $
\frac{1}{k} \sum_{i=1}^k I_i^{(1)} = \frac{1}{k} I_1 \lfloor \frac{k}{T} \rfloor \to \frac{I_1}{T}, \quad \text{ as } k \to \infty$. 
Thus, for $n$ large enough, the second term on the right in \eqref{eq:thm2_2} is identically zero. Thus, we only need to show the summability of the first term on the right in \eqref{eq:thm2_2}. 
Towards this end, note that the probability $\Prob_1 \left(\sup_{k \geq n } \big| \frac{1}{k} \sum_{i=1}^k \tilde{Z}_i^{(1)}\big| > \frac{\epsilon}{2T} \right)$
is decreasing with increasing $n$. Using this fact, we can write
\begin{equation*}
\begin{split}
 \sum_{n=1}^\infty \Prob_1 &\left(\sup_{k \geq n } \bigg| \frac{1}{k} \sum_{i=1}^k \tilde{Z}_i^{(1)}\bigg| > \frac{\epsilon}{2T} \right) \\
 &\leq T \sum_{n=1}^\infty \Prob_1 \left(\sup_{k \geq nT+1 } \bigg| \frac{1}{k} \sum_{i=1}^k \tilde{Z}_i^{(1)}\bigg| > \frac{\epsilon}{2T} \right).
\end{split}
\end{equation*}
Now, note that $\big| \sum_{i=1}^k \tilde{Z}_i^{(1)}\big|$ changes only when $k$ changes by multiples of $T$: $k=1, T+1, 2T+1, \cdots$. As a result, in between
these multiples of $T$, $\big| \frac{1}{k} \sum_{i=1}^k \tilde{Z}_i^{(1)}\big|$ monotonically decreases with increase in $k$. 
This implies that
\begin{equation*}
\begin{split}
 \sum_{n=1}^\infty& \Prob_1  \left(\sup_{k \geq nT+1 } \bigg| \frac{1}{k} \sum_{i=1}^k \tilde{Z}_i^{(1)}\bigg| > \frac{\epsilon}{2T} \right) \\
 &=\sum_{n=1}^\infty \Prob_1  \left(\sup_{k=mT+1; m \geq n} \bigg| \frac{1}{k} \sum_{i=1}^k \tilde{Z}_i^{(1)}\bigg| > \frac{\epsilon}{2T} \right) \\
 &=\sum_{n=1}^\infty \Prob_1  \left(\sup_{m \geq n} \bigg| \frac{1}{mT+1} \sum_{i=1}^{mT+1} \tilde{Z}_i^{(1)}\bigg| > \frac{\epsilon}{2T} \right) \\
 &\leq \sum_{n=1}^\infty \Prob_1  \left(\sup_{m \geq n} \bigg| \frac{1}{m} \sum_{i=1}^{mT+1} \tilde{Z}_i^{(1)}\bigg| > \frac{\epsilon}{2T} \right). 
\end{split}
\end{equation*}
The rightmost summation is finite if \eqref{eq:finitevarLLR} is satisfied because the sum inside the supremum $\sum_{i=1}^{mT+1} \tilde{Z}_i^{(1)}$ is a sum of $m$ i.i.d. random variables 
with distribution of $Z_1$ under $\Prob_1$ \cite{tart-niki-bass-2014}, \cite{tart-veer-siamtpa-2005}. The rest of the arguments follow from Theorem 3 in \cite{tart-veer-siamtpa-2005}. 
\end{IEEEproof}


\balance


\bibliographystyle{IEEEtran}
\bibliography{QCD_verSubmitted}

\end{document}